\documentclass[iop]{emulateapj}
\usepackage{apjfonts}

\usepackage{graphicx}
\usepackage{amsmath}
\usepackage{appendix}
\usepackage{amssymb}
\usepackage{aas_macros}
\usepackage{natbib}
\usepackage{latexsym}
\usepackage{longtable}
\usepackage{threeparttable}
\usepackage{subfigure}
\usepackage[figoff]{figcaps}
\usepackage{ccaption}
\usepackage{url}
\captiondelim{ }

\begin{document}

\pagestyle{plain}

\title{Architecture of Planetary Systems Based on {\em Kepler} Data:\\Number of Planets and Coplanarity}

\author{Julia Fang\altaffilmark{1} and Jean-Luc Margot\altaffilmark{1,2}}
\altaffiltext{1}{Department of Physics and Astronomy, University of California, Los Angeles, CA 90095, USA}
\altaffiltext{2}{Department of Earth and Space Sciences, University of California, Los Angeles, CA 90095, USA}

\begin{abstract} 

We investigated the underlying architecture of planetary systems by
deriving the distribution of planet multiplicity (number of planets)
and the distribution of orbital inclinations based on the sample of
planet candidates discovered by the {\em Kepler} mission.  The scope
of our study included solar-like stars and planets with orbital
periods less than 200 days and with radii between 1.5 and 30 Earth
radii, and was based on {\em Kepler} planet candidates detected during
Quarters 1 through 6. We created models of planetary systems with
different distributions of planet multiplicity and inclinations,
simulated observations of these systems by {\em Kepler}, and compared
the properties of the transits of detectable objects to actual {\em
Kepler} planet detections. Specifically, we compared with both the
{\em Kepler} sample's transit numbers and normalized transit duration
ratios in order to determine each model's goodness-of-fit. We did not
include any constraints from radial velocity surveys.  Based on our
best-fit models, 75-80\% of planetary systems have 1 or 2 planets with
orbital periods less than 200 days. In addition, over 85\% of planets
have orbital inclinations less than 3 degrees (relative to a common
reference plane). This high degree of coplanarity is comparable to
that seen in our Solar System.  These results have implications for
planet formation and evolution theories. Low inclinations are
consistent with planets forming in a protoplanetary disk, followed by
evolution without significant and lasting perturbations from other
bodies capable of increasing inclinations.

\end{abstract}
\keywords{methods: statistical -- planetary systems -- planets and satellites: general -- planets and satellites: detection}
\maketitle

\section{Introduction}

Knowledge of the architecture of planetary systems can provide
important constraints and insights into theories of planet formation
and evolution. For instance, planetary systems with a high degree of
coplanarity or alignment, such as the Solar System, are consistent
with the standard formation model of planets forming in a
protoplanetary disk. Additionally, planetary systems with inclined or
misaligned orbits can be indicative of past events that increased
eccentricities and inclinations (e.g., Kozai oscillations by outlying
perturbers, resonant encounters between planets, or planet-planet
scattering).  Consequently, information on planetary multiplicity and
the distribution of inclinations can reveal fundamental events in the
lives of planetary systems as well as test theories of planet
formation and evolution.

Because of observational biases, it may be difficult to reliably
assess the underlying multiplicity of planets and their inclination
distributions.  Planets may go undetected if they have masses/radii
below detection limits or if they are non-transiting (for the case of
transit surveys). As for inclinations, mutual (or relative)
inclinations between planets in the same system have only been
measured in a limited number of cases or indirectly
constrained. Examples include the planets orbiting the pulsar PSR
B1257+12 \citep{kona03}, the planets GJ 876 b and c using radial
velocity and/or astrometric data \citep{bean09,corr10,balu11},
$\upsilon$ And planets c and d using astrometric and radial velocity
data \citep{mcar10}, using transit timing/duration analyses to
constrain inclinations in the Kepler-9 system \citep{holm10},
Kepler-10 system \citep{bata11,fres11}, and Kepler-11 system
\citep{liss11_kep11}, and analyzing transits of planets over starspots 
for Kepler-30 \citep{sanc12}.

However, there now exists substantial knowledge about a large sample
of planetary systems that can be used to statistically determine the
underlying number of planets in each system and their relative orbital
inclinations.  In particular, the haul of planetary candidates
cataloged by the {\em Kepler} mission \citep[current count is over
  2,300; ][]{boru11a,boru11b,bata12} provides a large trove of systems
to study. These {\em Kepler} detections are termed planetary
candidates, because most of them have not been formally confirmed as real
planets by independent observations such as radial velocity detections
or successful elimination of false positive scenarios.  We will refer
to {\em Kepler} planetary candidates interchangeably as planetary
``candidates'' or ``planets.''

In the present study, we use the latest {\em Kepler} catalog of
planetary candidates (identified in Quarters 1 through 6) released by
\citet{bata12} to constrain the underlying multiplicity and
inclinations of planetary systems discovered by {\em Kepler}. By
``underlying,'' we are referring to our estimate of the true
distributions free of observational biases.  
In our analysis, we create thousands of model
populations, each obeying different underlying multiplicity and
inclination distributions. These planetary systems are subject to
artificial observations to determine which planets could be transiting
and detectable by {\em Kepler}. The detectable planets and their
properties are compared to the observed sample to determine each
model's goodness-of-fit to the data. Our results show that the
underlying architecture of planetary systems is typically thin (low
relative inclinations) with few planets with orbital periods under 200
days.  Similarly, other statistical studies have been performed to
constrain multiplicity and/or inclinations using different
methodologies or data sets
\citep[e.g.,][]{liss11,trem12,figu12,fabr12,weis12,joha12}. Our
analysis improves on previous work by including all available {\em Kepler} 
quarters, extending to 200-day orbital periods, and fitting models to
observables such as normalized transit duration ratios that contain
information on mutual orbital inclinations; these improvements lend to
a deeper investigation of the intrinsic distributions of planetary
systems.

This paper is organized as follows. Section \ref{meth_popsim} provides
details on how we created model populations including choice of
stellar and planetary properties. Section \ref{meth_evalsim} describes
how we determined detectable, transiting planets from our simulated
populations as well as our methods for determining each model's
goodness-of-fit compared to observations. Section \ref{results}
presents our main results with analysis of our best-fit 
models. Section \ref{discussion} includes comparisons of our results
with the Solar System and with previous work. Lastly, Section
\ref{conclusion} summarizes the main conclusions of our study.

\section{Creating Model Populations} \label{meth_popsim}

This section describes our methods for creating model
populations. Each model population consists of approximately 10$^6$
simulated planetary systems from which we determine planets detectable
by {\em Kepler}.  In total, we created thousands of model populations
obeying different underlying planet (1) multiplicity and (2)
inclination distributions.

\subsection{Stellar Properties} \label{meth_stars}

Each model population's planetary systems have host stars with
parameters drawn from a distribution of stellar properties. For each
simulated system, we randomly drew a star from a subset of the
quarterly KIC \citep[{\em Kepler} Input Catalog; ][]{brow11} 
CDPP (Combined Differential Photometric Precision) lists\footnote[1]{\url{http://archive.stsci.edu/pub/kepler/catalogs/}} 
\citep{chri12}.  There is a separate list per quarter,
because each file only lists targets observed that quarter and their
corresponding CDPP noise levels over 3, 6, and 12 hours. These
quarterly lists also include basic KIC parameters (unchanging in the
quarterly lists) for each observed star such as its {\em Kepler}
magnitude $K_p$, effective temperature $T_{\rm eff}$, surface gravity
parameter log($g$), and radius $R_*$.

We created a subset of the KIC CDPP lists in the following
manner. First, we collected the KIC CDPP lists for Quarters 1 to 6
(Q1-6). Not all stars are observed each quarter and we counted a total
of 189,998 unique stars observed at some point during Q1-6. Next, we
filtered this list of unique stars such that we only included stars that were 
observed under the ``exoplanet'' program, have nonzero values for
$R_*$ and CDPP entries, and obey the following restrictions:
\begin{gather}
	\nonumber           4100~{\rm K} \leq T_{\rm eff} \leq 6100~{\rm K}, \\
	\label{stellarcuts} 4.0 \leq {\rm log}(g\ [{\rm cm~s}^{-2}])\leq 4.9, \\ 
	\nonumber           K_p \leq 15~{\rm mag}. 
\end{gather} 
These stellar cuts allowed 
us to consider the brighter main-sequence stars with well-characterized 
properties in the KIC \citep[consistent with those in][]{howa11}.
With such filtering, we were left with a KIC subset of 59,224
remaining unique stars, which is roughly one-third of the original
list of unique stars. For each star in this KIC subset, we calculated
median CDPP values over 3, 6, and 12 hours from all the values
available in the Q1-6 lists.  We also used each star's surface gravity
parameter log($g$) and $R_*$ values, which are common across the Q1-6
KIC CDPP lists, to calculate each star's mass.  Finally, we recorded
each star's observing history over Q1-6 as indicated by the quarterly
KIC CDPP lists.

Our subset of the KIC with 59,224 stars was the catalog from which we
randomly picked stellar hosts (with their corresponding properties)
for each simulated planetary system. In the next subsections, we
describe how we chose physical and orbital parameters for the planets
in each system.

\subsection{Planet Period and Radius Distribution} \label{meth_periodradius}

We assigned values for the period and radius of each simulated planet
by drawing from debiased distributions.  To create such debiased
distributions, we started with the observed sample of {\em Kepler}
candidate planets (or {\em Kepler} Objects of Interest; KOIs) and
converted it to a debiased sample of planets using detection
efficiencies. First we will discuss how we filtered the KOI catalog,
and then we will discuss the calculation and application of detection
efficiencies.

Our observed sample of KOIs is based on the February 2012 release
(Q1-6) by \citet{bata12} consisting of 2,321 planetary candidates in
1,790 systems.  We only considered a subset of this KOI catalog by
applying the following filters. First, we only considered planet
candidates with positive values for their entries: we removed any
planet candidates with negative values of the orbital period because
such candidates are only based on a single transit \citep{bata12}, and
we also removed KOI 1611.02, the circumbinary planet also known as
Kepler-16b \citep{doyl11}, because many of its entries in the catalog
have negative values.  This cut left us with 2,298 remaining
candidates. Next, we filtered out any planetary candidates with host
stars that did not obey Equation (\ref{stellarcuts}) for consistency
with the simulated stellar population. This step filtered out about
half of the sample; 1,135 candidates remained. Lastly, we made
additional cuts based on planet radius, period, and SNR:
\begin{gather}
	\nonumber          P \leq 200~{\rm days}, \\
	\label{planetcuts} 1.5~R_{\earth} \leq R \leq 30~R_{\earth}, \\
	\nonumber          {\rm SNR (\rightarrow Q8)} \geq 11.5.
\end{gather}
These specific cuts were performed in order to choose a sample of
planetary candidates with period, radius, and SNR that are unlikely to
be missed by the {\em Kepler} detection pipeline. It is necessary to
choose a sample with a high degree of completeness because we want to
compare this sample to simulations, which are normally 100\% complete.  For
orbital periods, a range extending to 200 days ensures that multiple
transits are likely to have been observed during Q1-6, which covers
approximately 486.5 days.  For planet radii, the lower bound on $R$ of
$1.5~R_{\earth}$ was chosen because the sample of observed planets
with radii $R \lesssim 1.5~R_{\earth}$ has much lower detection
efficiencies \citep[e.g., ][]{howa11,youd11}.  For SNR, we required a
SNR$\geq$10 for Q1-6, which corresponds to SNR$\geq$11.5 for Q1-8 by
assuming that SNR roughly scales as $\sqrt{N}$, where $N$ is the
number of observed transits. This scaling is performed because the
SNRs of observed KOIs have been reported for Q1-8, not Q1-6, in
\citet{bata12}. To investigate incompleteness issues, we have also 
repeated all analyses in this paper for SNR cuts of 15, 20, and 25. 

Now we discuss how we calculated detection efficiencies, and how they
were used to debias this filtered sample of planetary candidates.
Following the formalism of \citet{youd11}, for any given planet the
net detection efficiency $\eta$ is the product of (1) the geometrical
selection effect $\eta_{\rm geom}$ and (2) the {\em Kepler}
photometric selection effect $\eta_{\rm phot}$.  The geometrical
selection effect is due to the probability of transit as planets with
longer periods or larger semi-major axes will have a lower probability
of crossing in front of the disk of the star as seen from {\em
  Kepler}, and is calculated as $\eta_{\rm geom}=R_*/a$, 
where $a$ is semi-major axis. The {\em
  Kepler} photometric selection effect is due to the photometric
quality of the star for detecting planets with specified radius and
period. Consequently, $\eta_{\rm phot}$ is a function of both
planetary radius and period. For a planet with radius $R$ and period
$P$, we calculated $\eta_{\rm phot}$ as the fraction of stars that can
detect such a planet with SNR$\geq$10 (see Section \ref{evalsim_snr}), based on the KIC sample
obtained in the previous section. Lastly, we calculated the net
detection efficiency as $\eta=\eta_{\rm geom}\eta_{\rm phot}$.

\begin{figure}[htb]
	\centering
	\includegraphics[width=3.2in]{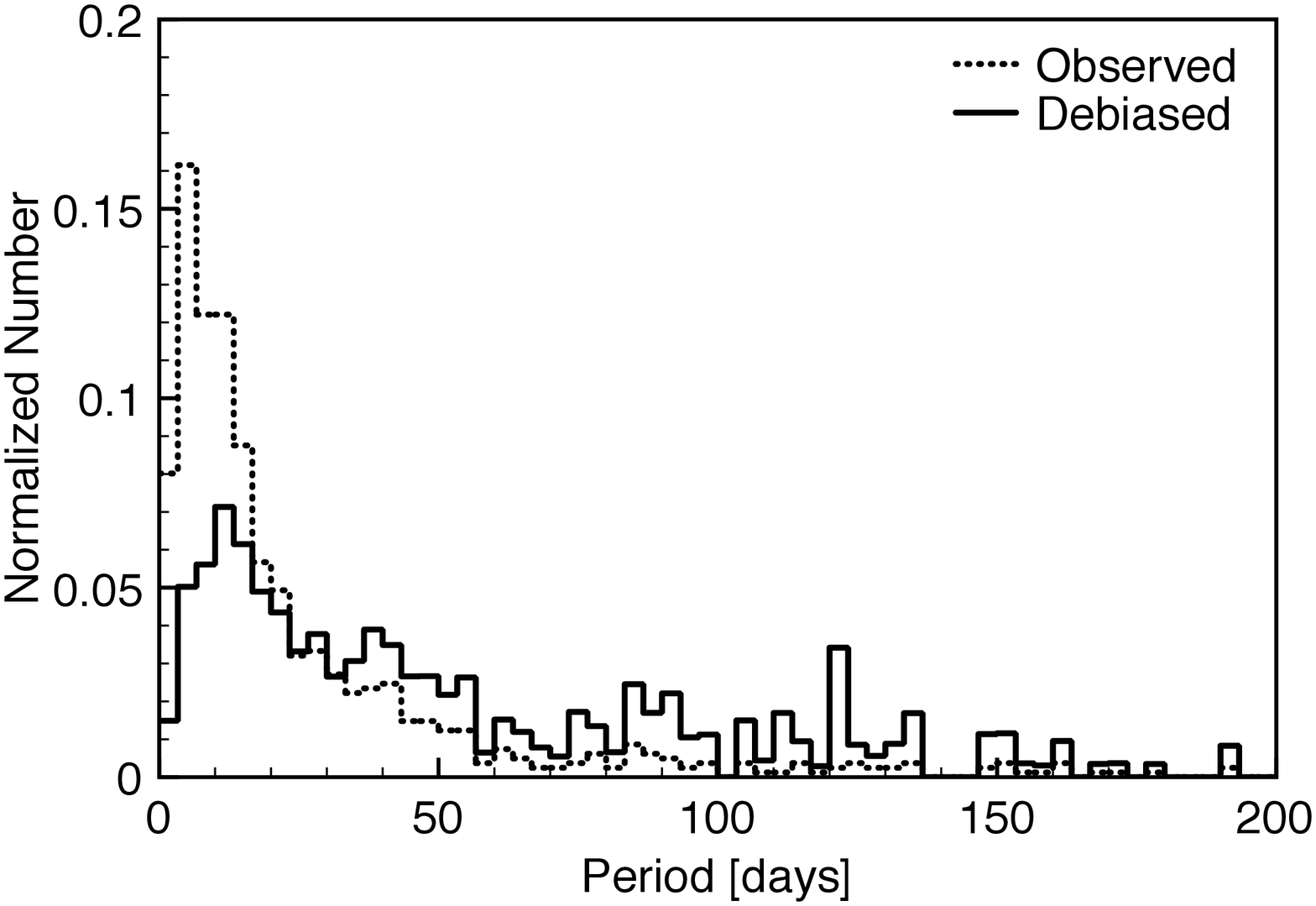}
	\includegraphics[width=3.2in]{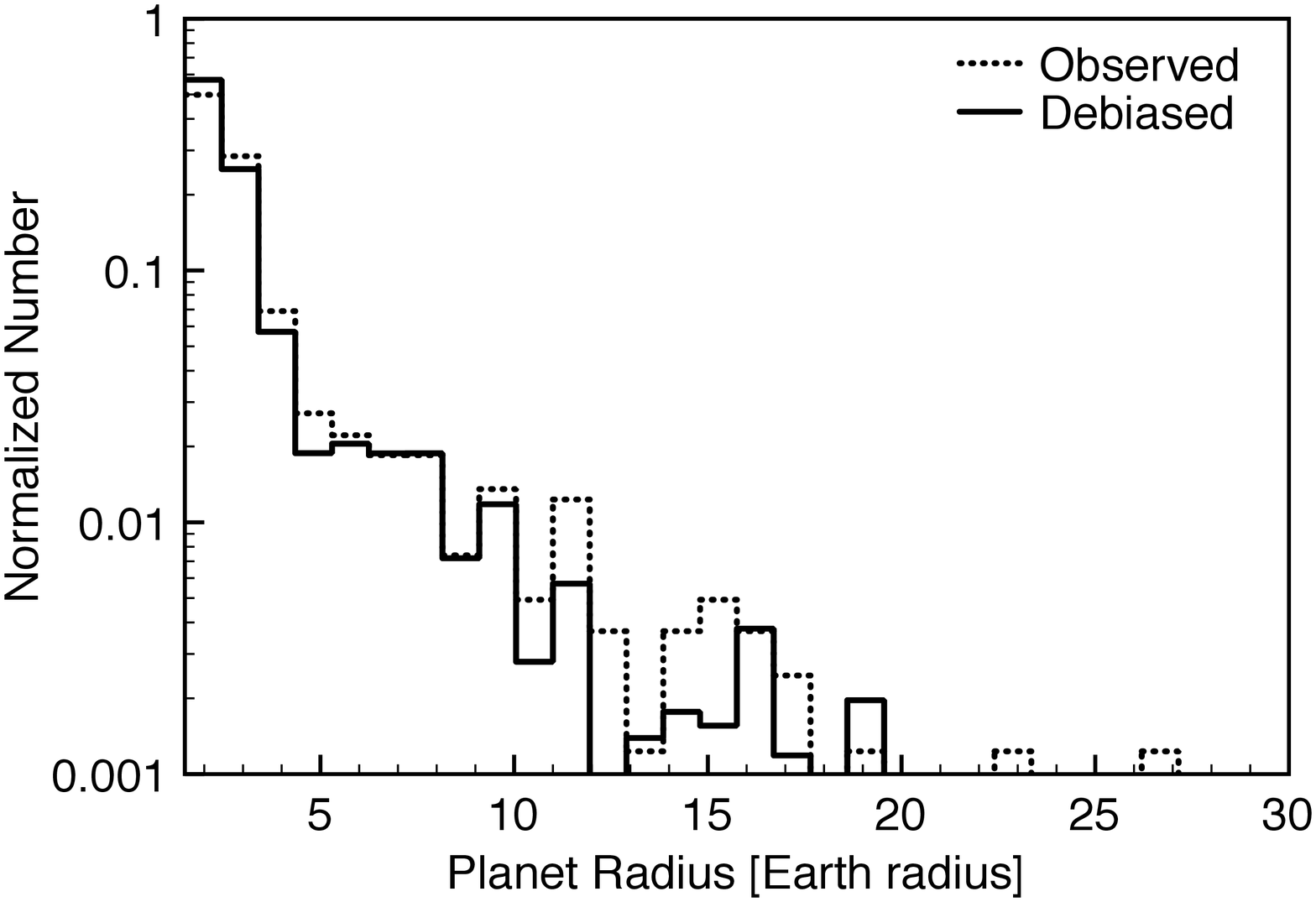}
	\caption{ Observed and debiased distributions of orbital
          periods and radii for a subset (Equation (\ref{planetcuts}))
          of the planets detected by {\em Kepler}.  The {\em top} plot
          for orbital period shows a distribution for 0$-$200 days
          divided into 60 equal-sized bins. The observed distribution
          shows increasing planet frequency at shorter periods, mainly
          due to the greater geometric probability of observing
          transiting planets at shorter periods. The {\em bottom} plot
          for planet radius shows a distribution for 1.5$-$30
          $R_{\Earth}$ divided into 30 equal-sized bins.
\label{debiased_dist}} 
\end{figure}

We applied the above methods for obtaining detection efficiencies to
our KOI subset of observed planetary candidates. For each candidate
planet (or detection) in our KOI subset, we calculated its net
efficiency $\eta$ given its observed period $P$ and radius $R$. Since
$\eta$ is the ratio of the number of detectable events to the number
of actual planets, the inverse of $\eta$ is equal to an estimate of
the actual number of planets represented by each detection. After we
calculated the value of $1/\eta$ for each detection in our KOI subset,
we created debiased, binned histograms for $P$ and $R$ representing an
estimate of the actual number of planets per $P$ or $R$ bin. The
normalized versions of these histograms are given in Figure
\ref{debiased_dist}.  We used these discrete distributions
representing debiased planet periods and radii in order to randomly
select values for each simulated planet's period and radius.

Our methods of obtaining photometric detection efficiencies
($\eta_{\rm phot}$) are the same as the methods described in
\citet{howa11} used to obtain binned (in $P$ and $R$) efficiencies
that were later analytically fit by \citet{youd11}, except for the
following differences.  First, we applied these methods to Q1-6 data
(previous work used Q2 only).  Second, we avoided any loss of
information that can occur during the binning process and/or the power
law fitting process by skipping those steps and directly calculating
the detection efficiency for each individual KOI candidate in our
filtered sample. Third, we calculated SNR differently by actually
using each KIC star's observing history during Q1-6 (e.g., not all
stars are observed each quarter) as well as some other minor
differences; our methods for calculating SNR are detailed in
Section \ref{evalsim_snr}.

\subsection{Planet Multiplicity Distribution} \label{meth_multiplicity}

We chose to represent the multiplicity distribution of simulated
planetary systems--the distribution of the number of planets per
star--by either a delta-like distribution, a modified Poisson
distribution, or a bounded uniform distribution.  All distributions
are described by a single parameter. We did not consider other
distributions such as an exponential distribution, which may result in
a bad fit to the data \citep{liss11}. All stars in our model
populations are assumed to have at least 1 planet, and so our study
cannot directly address the question of planet occurrence.

A delta-like distribution is a discrete distribution described by a
single parameter $D$. If the parameter $D$ represents an integer
value, than all planetary systems following this distribution will
have $D$ planets each. If the parameter $D$ does not represent an
integer value, then each planetary system can have either
$\lfloor$D$\rfloor$ (floor function) or $\lceil$D$\rceil$ (ceiling
function) planets.  For instance, a model population with $D=3$ means
that all of its planetary systems have 3 planets per star, and a model
population with $D=4.75$ means that 25\% of its planetary systems have
4 planets and 75\% of its planetary systems have 5 planets.  As a
result, $D$ also represents the mean number of planets per system.
This is the same as the ``uniform'' distribution used by \citet{liss11}.

A modified Poisson distribution described by parameter $\lambda$ is a
discrete distribution that is equivalent to a regular Poisson
distribution with mean value $\lambda$, except that we ignore zero
values. Therefore, by using a modified Poisson distribution we
restrict the regular Poisson distribution so that values picked from
the distribution must be greater than 0 (so that each planetary system
must have at least 1 planet).  As a result, the modified Poisson
distribution is not strictly Poisson since its mean value is different
from $\lambda$, hence the name ``modified Poisson.''  Mathematically,
the regular Poisson distribution can be written as
\begin{align} \label{poisson_equation}
	{\rm P}(k) = \dfrac{\lambda^k e^{-\lambda}}{k!},
\end{align} 
which gives the probability of obtaining a discrete value $k$ from
such a distribution.

A bounded uniform distribution is represented by a single parameter
$\lambda$. For each planetary system, first a modified Poisson
distribution (as defined above) with parameter $\lambda$ is used to
draw a value $N_{\rm max}$ (maximum number of planets). Second, a
discrete uniform distribution with range 1$-N_{\rm max}$ is used to
draw a value representing the number of planets in that system.

We varied the values of the parameter $D$ for the delta-like
distribution and the values of the parameter $\lambda$ for the
modified Poisson and bounded uniform distributions. In total, we
explored $D=2-7$ and $\lambda=1-6$, each with intervals of 0.25,
resulting in a total of 63 possibilities for the multiplicity
distribution.

\subsection{Planet Inclination Distribution}

We chose the inclination distribution in each planetary system to be
one of three possibilities: an aligned inclination distribution, a
Rayleigh distribution, or a Rayleigh of Rayleigh distribution, as also
used by \citet{liss11}.

An aligned distribution is the straightforward case where relative
inclinations in the system are 0$^{\circ}$. This is the same as a
perfectly coplanar system.

A Rayleigh distribution is a continuous distribution described by a
single parameter $\sigma$, which determines the mean and variance of
the distribution. Its mathematical form is
\begin{align} \label{rayleigh_equation}
	{\rm P}(k) = \dfrac{k}{\sigma^2}e^{-k^2/2\sigma^2},
\end{align}
which gives the probability density function for a value $k$ from this
distribution.

A Rayleigh of Rayleigh distribution, with a single parameter
$\sigma_{\sigma}$, means that we draw from two Rayleigh distributions
for each planetary system. First, the given parameter
$\sigma_{\sigma}$ defines the first Rayleigh distribution from which
we draw a value for $\sigma$ for each planetary system. Second, the
drawn value of $\sigma$ is then used to define the second Rayleigh
distribution from which we draw values for the inclinations of planets
in that system. This allows for the possibility that planetary systems
in a particular model population have Rayleigh inclination
distributions with different mean and variance.

We varied $\sigma$ for the Rayleigh distribution and $\sigma_{\sigma}$
for the Rayleigh of Rayleigh distribution with the same range of
values: 1$-$10$^{\circ}$ with intervals of 1$^{\circ}$, as well as 
values of 15$^{\circ}$, 20$^{\circ}$, and 30$^{\circ}$. In total, 
all possibilities for the 3 inclination distributions added up to a 
total of 27 possible inclination distributions.

\subsection{Other Planetary Parameters} \label{otherparams}

So far we have defined the distributions used in drawing the orbital
periods, radii, multiplicity, and orbital inclinations of simulated
planetary systems.  Now we discuss all other relevant planetary
parameters: planet mass, orbital semi-major axis, eccentricity, and
longitude of the ascending node. 

\begin{figure*}[htb]
	\centering
	\includegraphics[width=6.0in]{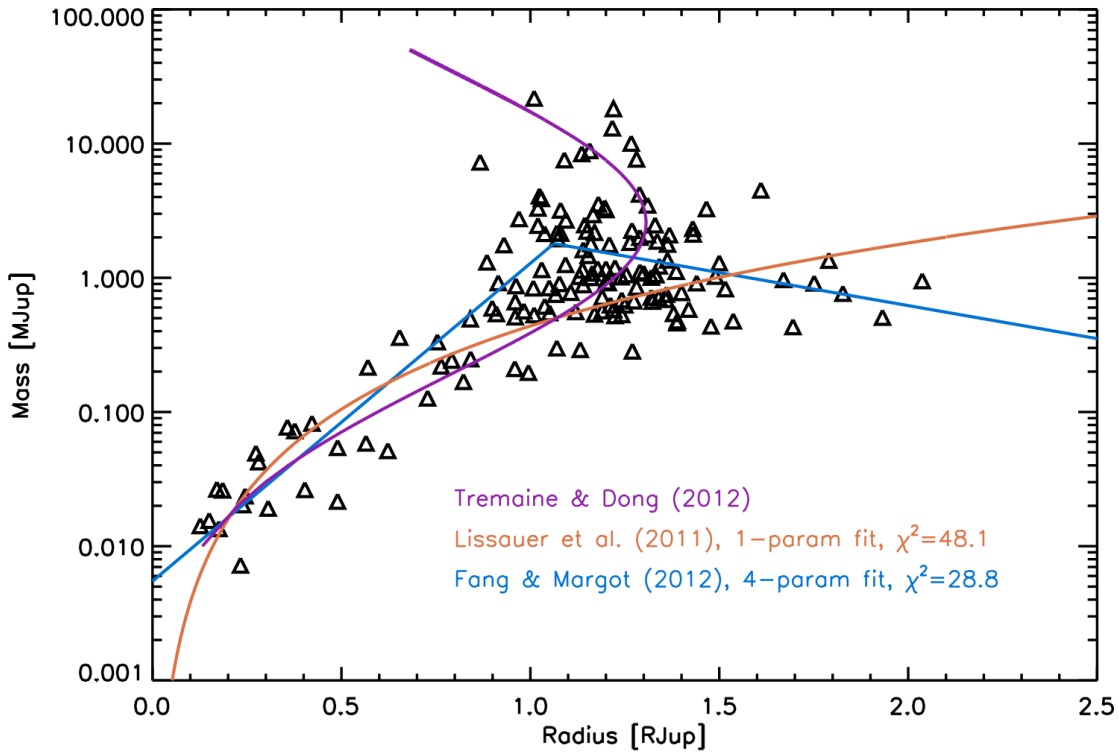}
	\caption{Known transiting planets (triangles) and three different mass-radius relations (colored curves). See Section \ref{otherparams} for details.
\label{m-r}} 
\end{figure*}

To determine planet mass $M$, we require an $M(R)$ relation to convert
from radius to mass. First, we used $(M/M_{\Earth}) =
(R/R_{\Earth})^{2.06}$, which is a commonly-used power law derived by
fitting to Earth and Saturn \citep{liss11}.  Results given in this
paper are based on this relation. Second, we obtained an alternate
$M(R)$ relation by fitting to masses and radii of known transiting
exoplanets. This $M(R)$ relation is a broken log-linear fit given as
\begin{align} \label{mrfit1}
	\log_{10} \left(\dfrac{M}{M_{\rm Jup}}\right) = 2.368 \left(\dfrac{R}{R_{\rm Jup}}\right) - 2.261 \\
	\nonumber {\rm for } \left(\dfrac{R}{R_{\rm Jup}}\right) < 1.062,
\end{align}
\begin{align} \label{mrfit2}
	\log_{10} \left(\dfrac{M}{M_{\rm Jup}}\right) = -0.492 \left(\dfrac{R}{R_{\rm Jup}}\right) + 0.777 \\
	\nonumber {\rm for } \left(\dfrac{R}{R_{\rm Jup}}\right) \geq 1.062.
\end{align}
This fit is plotted in Figure \ref{m-r}, along with data of known
transiting exoplanets, the $(M/M_{\Earth}) = (R/R_{\Earth})^{2.06}$
relation by \citet{liss11}, and a $R(M)$ log-quadratic fit by
\citet{trem12}.  
Our 4-parameter fit is a better match to the data than the 1-parameter
prescription of $(M/M_{\Earth}) = (R/R_{\Earth})^{2.06}$ (improvement
in $\chi^2$ from 48.1 to 28.8), and this is not simply due to the
addition of model parameters (F-test at 99.96\% confidence level).

In contrast to the fit by \citet{trem12}, our fit is an $M(R)$
function and not a $R(M)$ function and hence is useful for cases that
require assigning masses corresponding to particular radii (i.e., our
fit does not have an ambiguity where some values of $R$ have multiple
corresponding values of $M$, which makes it impossible to select a
unique mass for a given radius). In addition, our fit covers the full
range of parameter space in radii; the fit by \citet{trem12} has a
maximum radius value of about 1.3 Jupiter radii. We have repeated the
analysis described in this paper using our fitted $M(R)$ relation in
Equations (\ref{mrfit1})-(\ref{mrfit2}), and we find essentially the
same results as when we used $(M/M_{\Earth}) = (R/R_{\Earth})^{2.06}$.

Each planet's semi-major axis is determined using the orbital
period and Kepler's Third Law. For eccentricity, we assumed that all
planets have circular orbits. {\em Kepler} candidates do not show
evidence for large eccentricities \citep{moor11}, but we note that
even moderate eccentricities ($<$0.25) could affect transit durations
in a manner similar to that caused by differences in inclinations of
up to a few degrees.  For the longitude of the ascending node of each
planet's orbit, we drew an angle from a uniform distribution with
values between 0$^{\circ}$ and 360$^{\circ}$.

Now we have described all relevant stellar and planetary parameters to
build a fictitious planetary system. In the next subsection, we
describe stability constraints placed on these systems.

\subsection{Stability Requirements}

The first stability constraint is Hill stability, and we describe our
implementation using the following example. In the construction of
each planetary system, we add a first planet with its parameters
chosen according to the methods already described.  If the system's
assigned multiplicity is greater than one, we add a second planet. The
second planet is accepted only if it is Hill stable with the first
planet, which occurs when
\begin{align} \label{deltaeqn1}
	\Delta = \frac{a_2-a_1}{R_{H1,2}} \gtrsim 3.46,
\end{align}
where $a$ is the semi-major axis and $R_{H1,2}$ is the mutual Hill radius defined as
\begin{align} \label{deltaeqn2}
	R_{H1,2} = \left( \frac{M_1+M_2}{3M_*} \right)^{1/3}\frac{a_1+a_2}{2},
\end{align}
\citep[e.g.,][]{glad93,cham96}.  Here, $M$ represents mass and
subscripts $1$ and $2$ denote the inner and outer planets,
respectively. If the second planet is not accepted because it violates
Hill stability with the first planet, then the second planet has its
radius and orbital period (or equivalently, mass and semi-major axis)
re-drawn until Hill stability is satisfied. We require that all
adjacent planet pairs satisfy this Hill stability requirement.  The
assumption that adjacent planet pairs should be Hill stable is
reasonable; out of the 885 planet candidates located in multi-planet
systems detected during Q1-6 of the {\em Kepler} mission, only two
pairs of planets are found to be apparently unstable by \citet{fabr12}
on the basis of numerical integrations using nominal mass-radius
scalings and circular orbit assumptions.

Following the methods of \citet{liss11}, we imposed a second stability
constraint that any adjacent planet trio (any three neighboring
planets) must satisfy 
\begin{align} \label{deltaeqn3}
	\Delta_{\rm inner~pair~of~trio} + \Delta_{\rm outer~pair~of~trio} \geq 18.
\end{align}
As described for the example above for Hill stability cases, if the
constraint in Equation (\ref{deltaeqn3}) is not satisfied, a 
planet's properties are re-drawn until this stability requirement is 
fulfilled.

For the creation of multi-planet systems, we added planets to each
system one at a time, where successful addition of a planet occurred
when it fulfilled both stability criteria described above. For systems
with a large multiplicity, at times it was not possible to satisfy
stability constraints given our period range and therefore we allowed
a maximum of 1000 attempts per star before discarding it. Therefore,
in rare cases, some model populations have a slightly lower number of
planetary systems.

\subsection{Summary}

To summarize this section: we have described the steps taken to
assemble various model populations, each consisting of about 10$^6$
planetary systems. Each of these model populations follows different
combinations of (1) multiplicity and (2) inclination
distributions. Taking all combinations of the values for the tunable
parameters in these distributions, there are a total of 1,701
possibilities. Our goal is to take these model populations, run them
through synthetic observations, and see which planets are transiting
and detectable. The properties of these detectable planets will then
be compared to the actual observed {\em Kepler} planets to determine
which model produces acceptable fits. This allows us to determine the
nature of underlying multiplicities and inclinations. In the next
section, we discuss how we evaluate these model populations.

\section{Evaluating Model Populations} \label{meth_evalsim}

We describe our methods for evaluating each model population: (1)
determining detectable, transiting planets, and (2) comparing them to
the observed {\em Kepler} planets using statistical tests.

\subsection{Determining Detectable, Transiting Planets} \label{evalsim_snr}

For each model population, we determined which of the planets in each
planetary system were {\em transiting}. From the group of transiting
planets, we determined which transiting planets were {\em detectable}
based on SNR requirements. The following paragraphs describe these
methods.

For each planetary system, we evaluated whether any of the planets were 
transiting.  First we picked a random line-of-sight (i.e., picking a
random point on the celestial sphere) from the observer to the system,
and the plane normal to this line-of-sight vector determines the plane
of sky.  We then computed the planet-star distance projected on the
plane of sky \citep[e.g., see ][]{winn2010}.  The minimum in that
quantity is the {\em impact parameter}, which we normalized by the
stellar radius.  A planet transits when its impact parameter is less
than 1.

We calculated the SNR of transit events in order to determine whether
or not each transiting planet would have been detectable during Q1-6
of the {\em Kepler} mission.  To do so, first we randomly picked a MJD
(Modified Julian Date) between the first and last cadence mid-times of
Q1-6 by using the dates provided in the {\em Kepler} Data Release
Notes\footnote[2]{The DRN are available for download from MAST at
  \url{http://archive.stsci.edu/kepler/data\_release.html}.}  (DRN) 14
and 16. We assumed that the planet transited on this randomly-picked
date, and we explored backwards and forwards in time using the
planet's orbital period in order to determine the dates of all
transits during Q1-6. Next, we eliminated any transits that occurred
in the gaps between quarters; such gaps are necessary to roll the
spacecraft 90$^{\circ}$ to keep its solar arrays illuminated. In
addition, not all stars are observed each quarter, for instance, due
to CCD module failure. As a result, we used the planet's host star
observing log that recorded all observed quarters (see Section
\ref{meth_stars}), and eliminated any transits that occurred in
quarters where the host star was not observed. Next, we eliminated a
small fraction of remaining transits in a random fashion to account
for a 95\% duty cycle because of observing downtimes (e.g., breaks for
data downlink).  After accounting for these issues, all remaining $N$
transits for this planet were then counted and used to calculate the
detection SNR as
\begin{align} \label{snr_equation}
	{\rm SNR} = \left(\dfrac{R}{R_*}\right)^2 \dfrac{\sqrt N}{\sigma_{\rm CDPP}},
\end{align}
where the first fraction represents the depth of the transit and
$\sigma_{\rm CDPP}$ represents an adimensional noise metric (quadratically interpolated
to the transit duration length using the set of measured \{3-hr, 6-hr,
12-hr\} CDPP values) for the planet's host star (see Section
\ref{meth_stars}). If the SNR is at least the value of the SNR
threshold for detection (SNR$\geq$10), then the transiting planet is
labeled as detectable.

\subsection{Comparing to Observations} \label{prob_methods}

Here we describe how we evaluated each model population's
goodness-of-fit when compared to observations. All model populations
are distinct from each other due to the different parameter values
chosen for their underlying multiplicity and inclination
distributions. To evaluate how well each model population matches the
observations, we compared them using 2 observables: (1) observed
multiplicity vector and (2) observed distribution of normalized,
transit duration ratios. These observables were computed for both the
simulated planets and the KOI candidates.

We discuss these 2 types of observables in greater detail. The
multiplicity vector $\boldsymbol\mu$ represents the number of systems
observed with $j$ transiting planets, where $j=1,2,3,4,5,6,7+$.  In
other words, the multiplicity vector describes how many systems are
observed to have a single planet, how many systems are observed with 2
planets, etc.  Systems with 7 or more planets are placed in the same
multiplicity category: $j=7+$. The normalized, transit duration ratio
$\xi$ is defined as \citep[e.g.,][]{fabr12}
\begin{align} \label{xieqn}
	\xi = \dfrac{T_{\rm dur,1}/P_1^{1/3}}{T_{\rm dur,2}/P_2^{1/3}},
\end{align}
where the transit duration $T_{\rm dur}$ is normalized by the orbital
period raised to the 1/3 power $P^{1/3}$. Subscripts $1$ and $2$
represent any pair of planets in the same system where $1$ denotes the
inner planet 
and $2$ denotes the outer planet.  A $j-$planet system has a total of
$j(j-1)/2$ pairs. We decided to fit to values of $\xi$ because they
contain information on mutual inclinations of observed multi-planet
systems. For instance, coplanar systems tend to have values of $\xi >
1$, because the inner planet's impact parameter is smaller and hence
its normalized transit duration is longer.
\citet{fabr12} used the observed {\em Kepler} distribution
of $\xi$ to determine that the observed mutual inclinations are low at
1.0-2.3$^{\circ}$.

We calculated the ($\boldsymbol\mu$, $\xi$) observables for the subset
of KOIs previously obtained in Section \ref{meth_periodradius}.  We
also calculated the ($\boldsymbol\mu$, $\xi$) observables for each
model population by considering all detectable, transiting planets.

Our next step was to obtain a statistical measure of each model
population's goodness-of-fit to the observed population by calculating
a test statistic and its corresponding probability. A larger
probability indicates a better match to observations. To compare
multiplicity vectors $\boldsymbol\mu$, we performed chi-square tests,
which are appropriate for discrete data.  Chi-square tests are valid
when the chi-square probability distribution is a good approximation
to the distribution of the chi-square statistic (Equation
(\ref{x2_stat})). As shown by theoretical investigations, the
approximation is usually good when there are at least 5 discrete
categories and at least 5 values in each category for the model
population's $\boldsymbol\mu$ \citep[e.g.,][]{Hoel84}.  These
requirements may not always be satisfied in our case, and so we also
perform a randomization method to calculate chi-square probabilities,
as described below.  To compare $\xi$ distributions, we performed
Kolmogorov-Smirnov (K-S) tests, which can be used for continuous
data. K-S tests have no restrictions on sample sizes, and are also
distribution-free and so therefore can be applied for any kind of
underlying distribution of the comparison samples
\citep[e.g.,][]{Stuart1991}.  These tests are described in more detail
below.

The chi-square test used for comparing multiplicity vectors uses a
chi-square statistic $\chi^2$ computed as \citep[e.g.,][]{pres92}
\begin{align} \label{x2_stat}
	\chi^2 = \sum_{j} \dfrac{(O_j - E_j)^2}{E_j},
\end{align}
where $O_j$ represents an observed quantity, in this case the number 
of observed planetary systems with $j$ planets, $E_j$ represents an
expected or theoretical value, in this case the number of simulated 
planetary systems with $j$ detectable planets, and $j$ represents the index of the
multiplicity vector (i.e., 1 for 1-planet systems, N for N-planet
systems, etc.) being summed over.  Scaling was performed to adjust the
model multiplicity vector such that $\sum O_j = \sum E_j$.  Large
values of $\chi^2$ represent worse fits and larger deviations between
$O_j$ and $E_j$ quantities, meaning it is not likely the $O_j$ values
are drawn from the population represented by the $E_j$ values. For
comparison between each model population's expected values and
observed values, we only considered $j$ indices where $E_j \not= 0$
(noting the appearance of $E_j$ in the denominator of Equation
(\ref{x2_stat})). Since we must ignore categories with
$E_j=0$, we note that there may be cases (i.e., for the delta-like
multiplicity distribution) where a given model population with only
1-planet and 2-planet systems may yield a low $\chi^2$ even though it
is in fact {\em not} a good match to the data, e.g., because it does not
predict any 3+ planet systems that are actually observed. 
Model populations based on delta-like multiplicity distributions do not 
provide a good match to the data (Section \ref{results}). Consequently, 
we will not consider them further and none of the $\chi^2$
probabilities presented in this paper suffer from this issue.

We computed two chi-square probabilities for each calculation of the
statistic in Equation (\ref{x2_stat}).  First, the standard chi-square
probability is an incomplete gamma function that can be computed with
knowledge of $\chi^2$ and the number of degrees of freedom. For our
case, the number of degrees of freedom is equal to the number of
indices with $E_j \not= 0$ subtracted by 1 (to account for the scaling
constraint).  This standard probability function is a good
approximation as long as the number of bins or the number of values in
each bin is large. Given that some higher-$j$ categories in our
multiplicity vector may have low numbers, we performed a second
calculation of the chi-square probability using a more robust
randomization method \citep[e.g.,][]{good06}. In this method, we
determined how often we could pick a new distribution from the model
distribution with $\chi^2_{\rm new,model}$ (comparing new and model
distributions) that was worse than the $\chi^2_{\rm obs,model}$
(comparing observed and model distributions).  In practice, we
randomly picked 10,000 times per model population and the fraction of
them that yielded $\chi^2_{\rm new,model}>\chi^2_{\rm obs,model}$
represented the chi-square probability.  We found that in most cases,
there was reasonable agreement between the chi-square probabilities
from the standard function and from the randomization method. The
chi-square probabilities quoted later in this paper are those obtained
from the randomization method.

For the K-S test used for comparing $\xi$ distributions, we calculated
the K-S statistic and the standard K-S probability
\citep[e.g.,][]{pres92}. The K-S statistic $D$ is the maximum
difference of the absolute value of two distributions given as
cumulative distribution functions, or
\begin{align} \label{ks_stat}
	D = {\rm max}|O_{N_1}(\xi) - E_{N_2}(\xi)|,
\end{align}
where $O_{N_1}(\xi)$ represents the observed cumulative distribution
function of $\xi$ values, $E_{N_2}(\xi)$ represents the expected or
model cumulative distribution function, and $N_1$ and $N_2$ is the
number of $\xi$ values for the observed and expected distributions,
respectively.  The significance or probability of $D$ represents the
chance of obtaining a higher value of $D$, and can be calculated with
knowledge of $D$ and the number of points $N_1$ and $N_2$.  
This test is valid even with an unequal number of points,
i.e., $N_1 \not= N_2$.  We note that $N_2$ can vary significantly
between model populations and this can affect the reliability of the
calculated probability, such as when comparing a probability obtained
from one model population to the probability obtained from another
model. This is most pertinent for cases with large differences in
$N_2$ (i.e., between model populations with small and large
multiplicities), where a poor fit is harder to discern when there are
fewer model values $N_2$ to compare to observed values.  This issue
can manifest itself as outliers in our results for model populations
with fewer number of planets per system. As we will show, we do not
find any evidence of such outliers in our results.

In the next section, we describe our calculations of these probabilities to
evaluate the match between each model population and the data.

\section{Results} \label{results}

\begin{figure*}[p]
	\centering
	\includegraphics[width=6.3in]{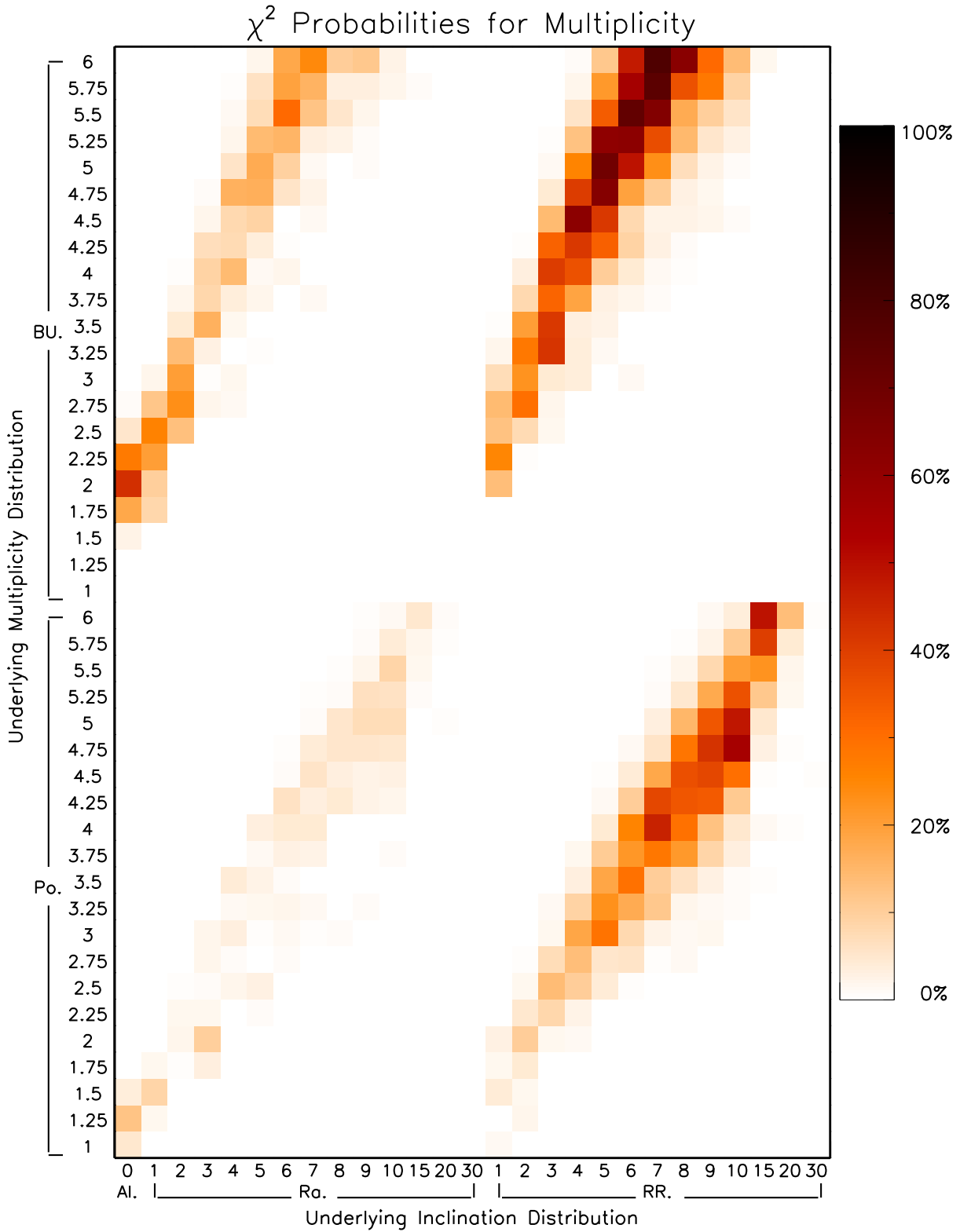}
	\caption{Diagram showing probabilities from $\chi^2$ tests
          that evaluated the goodness-of-fit of various models with
          respect to planet multiplicity.  Each square on this grid
          represents a different model, and each square's color
          depicts the probability (the higher the probability, the
          better the fit). The x-axis shows the {\em underlying
            inclination distribution}, and can be aligned (Al.),
          Rayleigh (Ra.), or Rayleigh of Rayleigh (RR.). For Al., the
          number given is the inclination, or 0$^{\circ}$. For Ra. and
          RR., the numbers indicate the $\sigma$ or $\sigma_{\sigma}$
          parameter (in degrees) for the distribution. The y-axis
          shows the {\em underlying multiplicity distribution}, and
          can be modified Poisson (Po.) or bounded uniform (BU.).  For
          Po, the number given is $\lambda$ of the Poisson
          distribution and represents approximately the mean number of
          planets per system. For BU., the number given is $\lambda$
          of the Poisson distribution from which the maximum number of
          planets is drawn. This plot can be approximately divided
          into 4 quadrants, where in each quadrant the y-axis has the
          number of planets per system increasing to the top and the
          x-axis has inclinations increasing to the right.
\label{statsmap_1}} 
\end{figure*}

\begin{figure*}[p]
	\centering
	\includegraphics[width=6.3in]{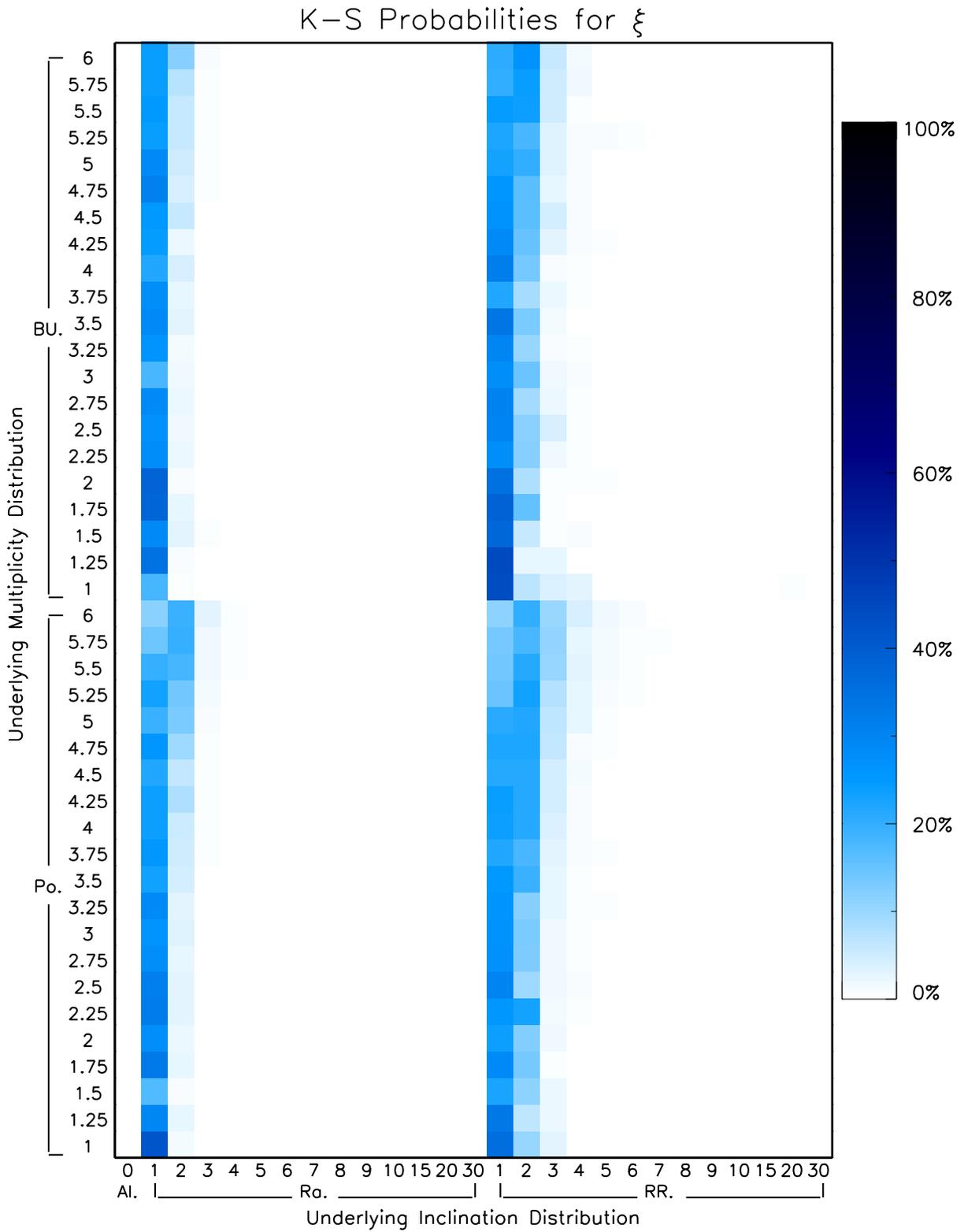}
	\caption{Same as Figure \ref{statsmap_1}, except this diagram shows K-S probabilities of $\xi$ distributions. These probabilities represent the goodness-of-fit of various models with respect to orbital inclinations.
\label{statsmap_2}} 
\end{figure*}

\begin{figure*}[p]
	\centering
	\includegraphics[width=6.3in]{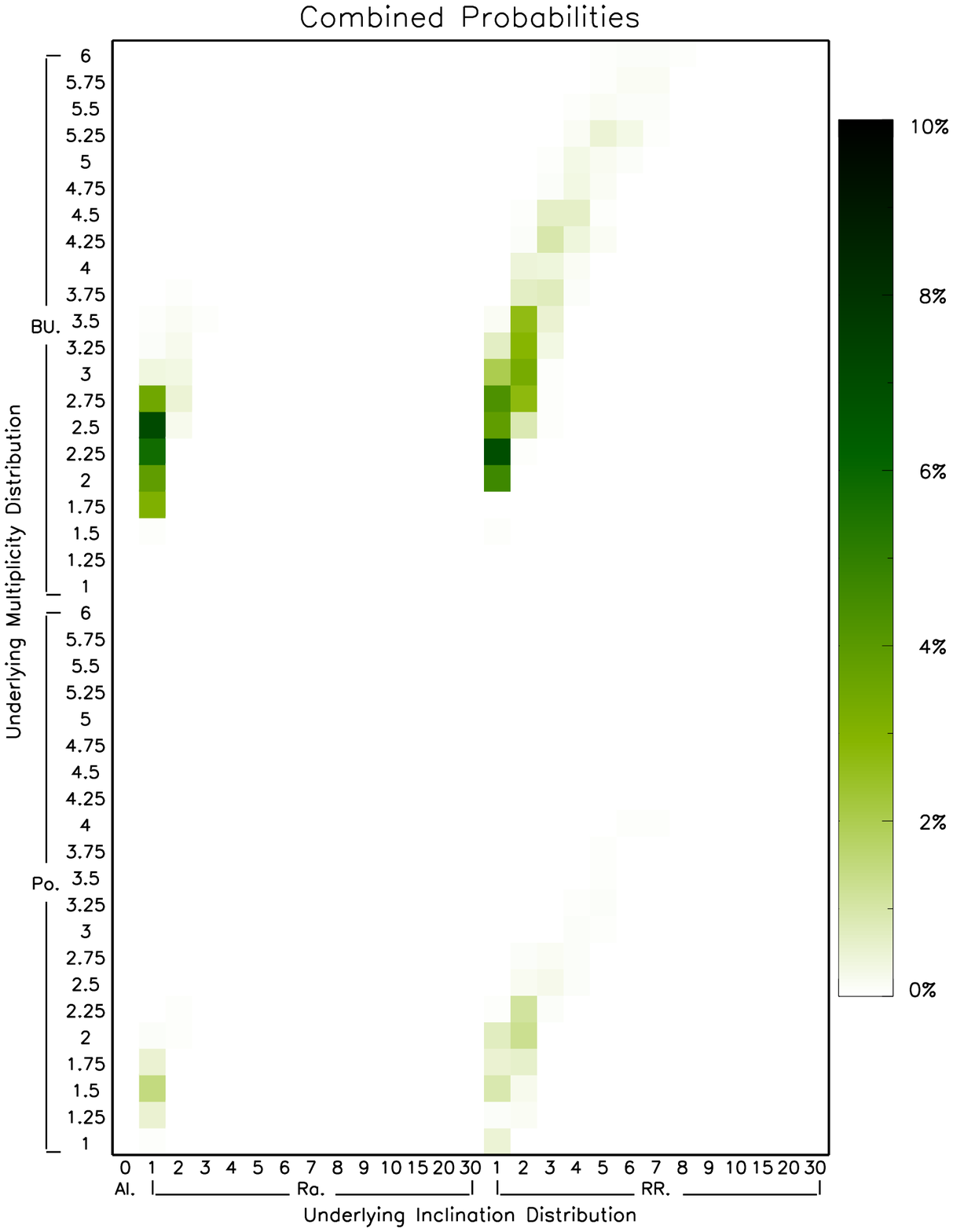}
	\caption{Same as Figure \ref{statsmap_1}, except this diagram shows combined (multiplied) probabilities from $\chi^2$ and K-S tests. These probabilities represent the overall goodness-of-fit of various models to the {\em Kepler} sample.
\label{statsmap_3}} 
\end{figure*}

In this section, we present the results for the fit between each model
population and the KOI subset that obeys Equations (\ref{stellarcuts})
and (\ref{planetcuts}), and these results are illustrated in Figures
\ref{statsmap_1}$-$\ref{statsmap_3}.  We do not show any results
incorporating the delta-like multiplicity distribution because it does 
not provide a good match to the data. We will describe each
of these figures in turn.

Figure \ref{statsmap_1} shows the probabilities resulting from the
$\chi^2$ test that compared the observed and model multiplicity
vectors (number of 1-planet systems, number of 2-planet systems,
etc.). Most models can be ruled out with 3$-$sigma confidence, and
these tend to be systems with many planets and low inclinations or
systems with few planets and high inclinations. On the other hand,
models that are consistent with the data include systems of few
planets with low inclinations or systems of many planets with high
inclinations; this degeneracy is evident in the upward slopes in
Figure \ref{statsmap_1}.  This degeneracy or upward trend is seen
because the intrinsic populations corresponding to the observed
transits can take a variety of forms: from few planets that are
well-aligned to many planets with larger inclinations.  The degeneracy
exists because systems with many planets and large inclinations are
detected by transit surveys as systems with few planets and small
inclinations.  Models with a bounded uniform multiplicity distribution
and a Rayleigh of Rayleigh inclination distribution appear to provide
a better fit (higher $\chi^2$ probabilities) than other models.

Figure \ref{statsmap_2} depicts the probabilities from the K-S test
that evaluated the fit between the observed and model $\xi$ (Equation
(\ref{xieqn})) distributions of multi-planet systems. The distribution
of $\xi$ values is used to gauge the extent of coplanarity or
non-coplanarity in multi-planet systems. This diagram shows that
higher probabilities are obtained for models with low-inclination
distributions. Nearly all models with a Rayleigh $\sigma$ or Rayleigh
of Rayleigh $\sigma_{\sigma}$ parameter greater than 3$^{\circ}$ can
be ruled out with 3$-$sigma confidence.  None of the models with
perfectly aligned (coplanar) systems provide acceptable fits.  The
best fits with highest probabilities are Rayleigh or Rayleigh of
Rayleigh inclination distributions with
$\sigma,\sigma_{\sigma}=1^{\circ}$.  The probabilities in this figure
appear to be relatively independent of the underlying multiplicity
distribution (evident in the long vertical columns of blue colors),
because the $\xi$ distribution probes the relative inclination between
any pair of planets in a system and is independent of the multiplicity
or number of planets. As a result, we find that the K-S test is not
very sensitive to the underlying multiplicity model.  Comparison
between Figures \ref{statsmap_1} and \ref{statsmap_2} shows that the
degeneracy evident in Figure \ref{statsmap_1} is broken when
considering Figure \ref{statsmap_2} as well--by also examining Figure
\ref{statsmap_2}, we see that only systems with fewer planets and
lower inclinations are consistent with the data.

Figure \ref{statsmap_3} illustrates the combined probabilities from
Figures \ref{statsmap_1} and \ref{statsmap_2}. The combined
probability is the product of the $\chi^2$ and K-S probabilities. We
make the assumption that these two probabilities are independent.
This product of two probabilities is shown in Figure \ref{statsmap_3},
from which it is evident that most models can be ruled out as
unacceptable fits. Degeneracies present in both Figures
\ref{statsmap_1} and \ref{statsmap_2} are broken when the data in
these figures are combined in Figure \ref{statsmap_3}. The best fits
are models with lower inclinations and lower multiplicities, and in
particular, the fits with a bounded uniform distribution in
multiplicity and a Rayleigh or Rayleigh of Rayleigh distribution in
inclination are most consistent with the data.

\subsection{Overall Best-Fit Models:\\Few Planets and Low Inclinations} \label{bestfit_section}

As seen in Figure \ref{statsmap_3}, the overall best-fit models with
the largest combined probabilities are models with low-multiplicity
bound uniform distributions (represented by $\lambda$) and with
low-inclination Rayleigh (represented by $\sigma$) or Rayleigh of
Rayleigh (represented by $\sigma_{\sigma}$) distributions.  The
best-fit model with a bounded uniform and Rayleigh distribution has
$\lambda=2.50$ and $\sigma=1^{\circ}$.  The best-fit model with a
bounded uniform and Rayleigh of Rayleigh distribution has
$\lambda=2.25$ and $\sigma_{\sigma}=1^{\circ}$. There are no
qualitative differences between the quality of these two fits.  Based
on these two best-fit models, we find that 75-80\% of planetary
systems have 1 or 2 planets with orbital periods less than 200
days. In addition, over 85\% of planets have orbital inclinations less
than 3 degrees (relative to a common reference plane).  These results 
represent our best estimate of the
underlying distributions of {\em Kepler} transiting systems for the
parameter space explored here (Equations (\ref{stellarcuts}) and
(\ref{planetcuts})).  Assuming that the {\em Kepler} sample is
representative, these distributions also describe the architecture of
planetary systems in general.

\begin{figure*}[thb]
	\centering 
	\vspace{-0.8cm}
	\mbox{ \subfigure{\includegraphics[width=2.8in]{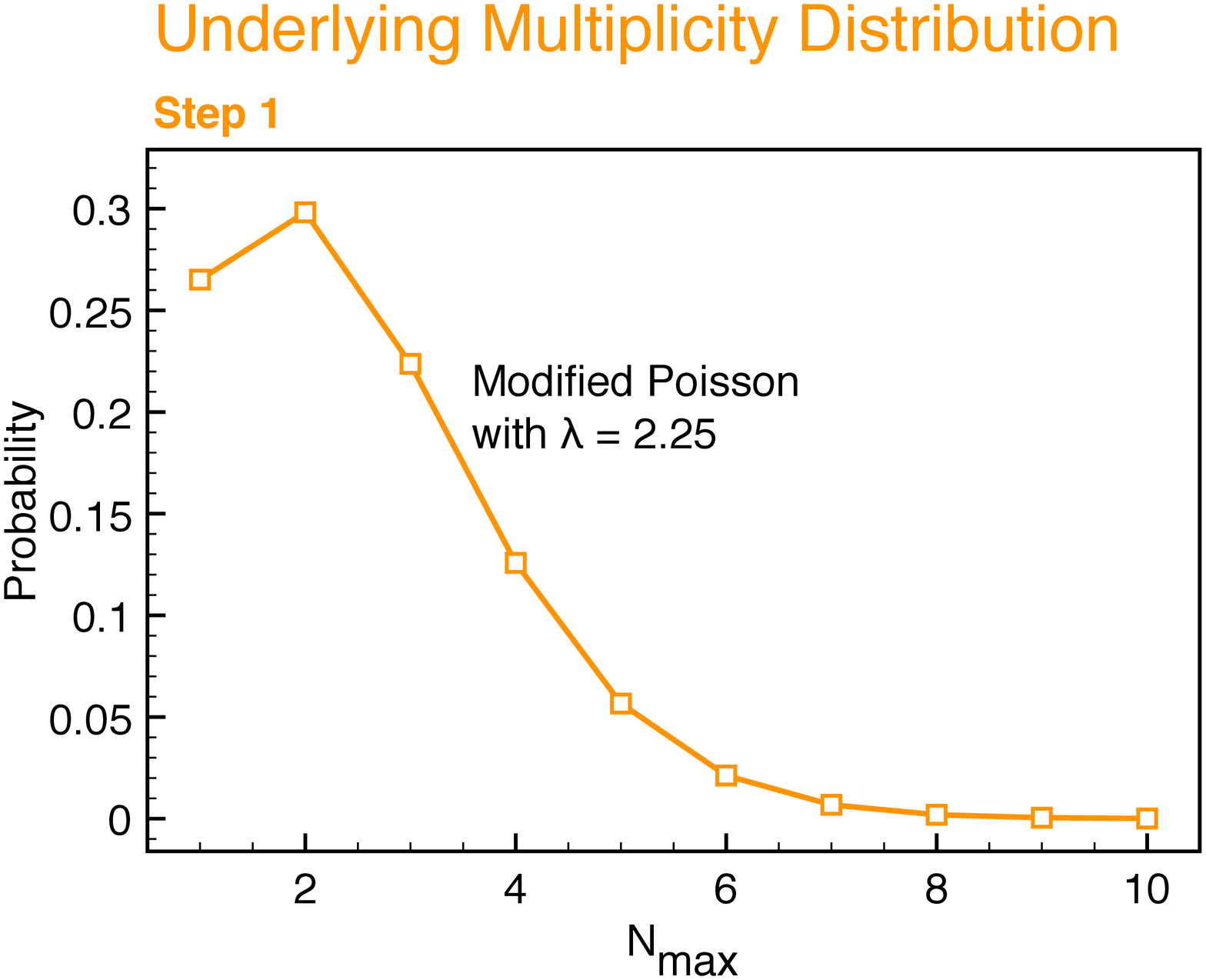}} \quad
	       \subfigure{\includegraphics[width=2.8in]{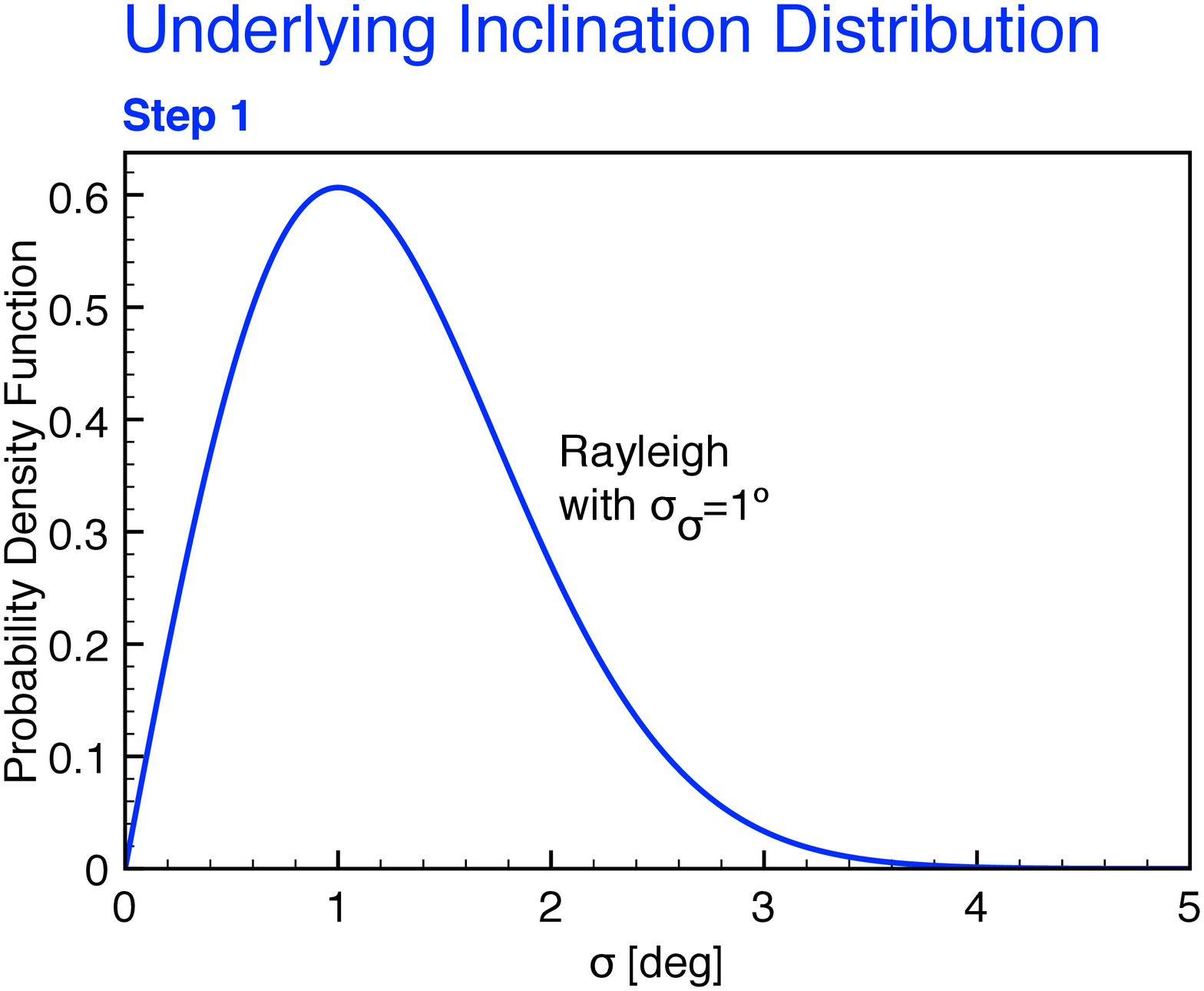}} }
	\vspace{-0.3cm}
	\mbox{ \subfigure{\includegraphics[width=2.8in]{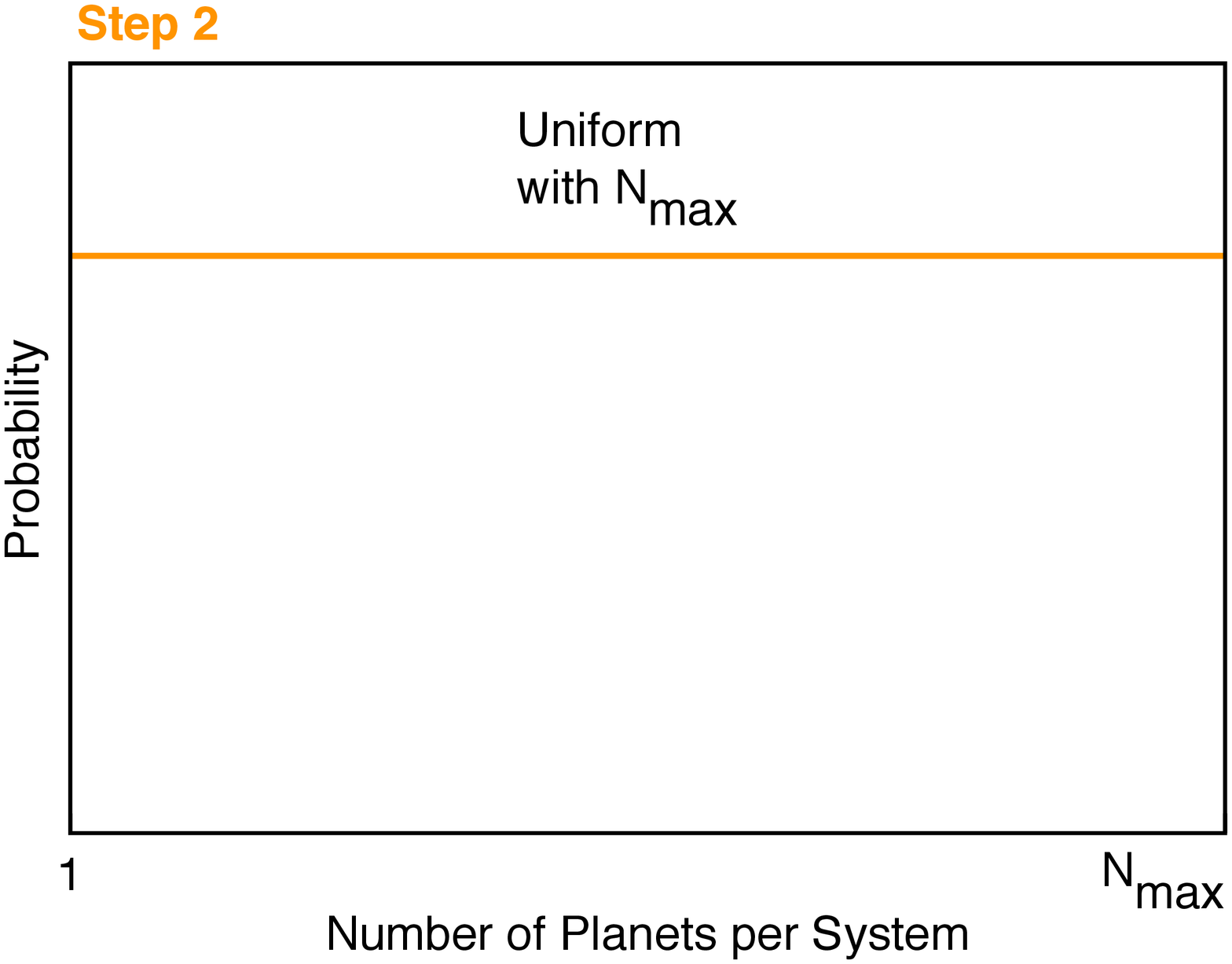}} \quad
	       \subfigure{\includegraphics[width=2.8in]{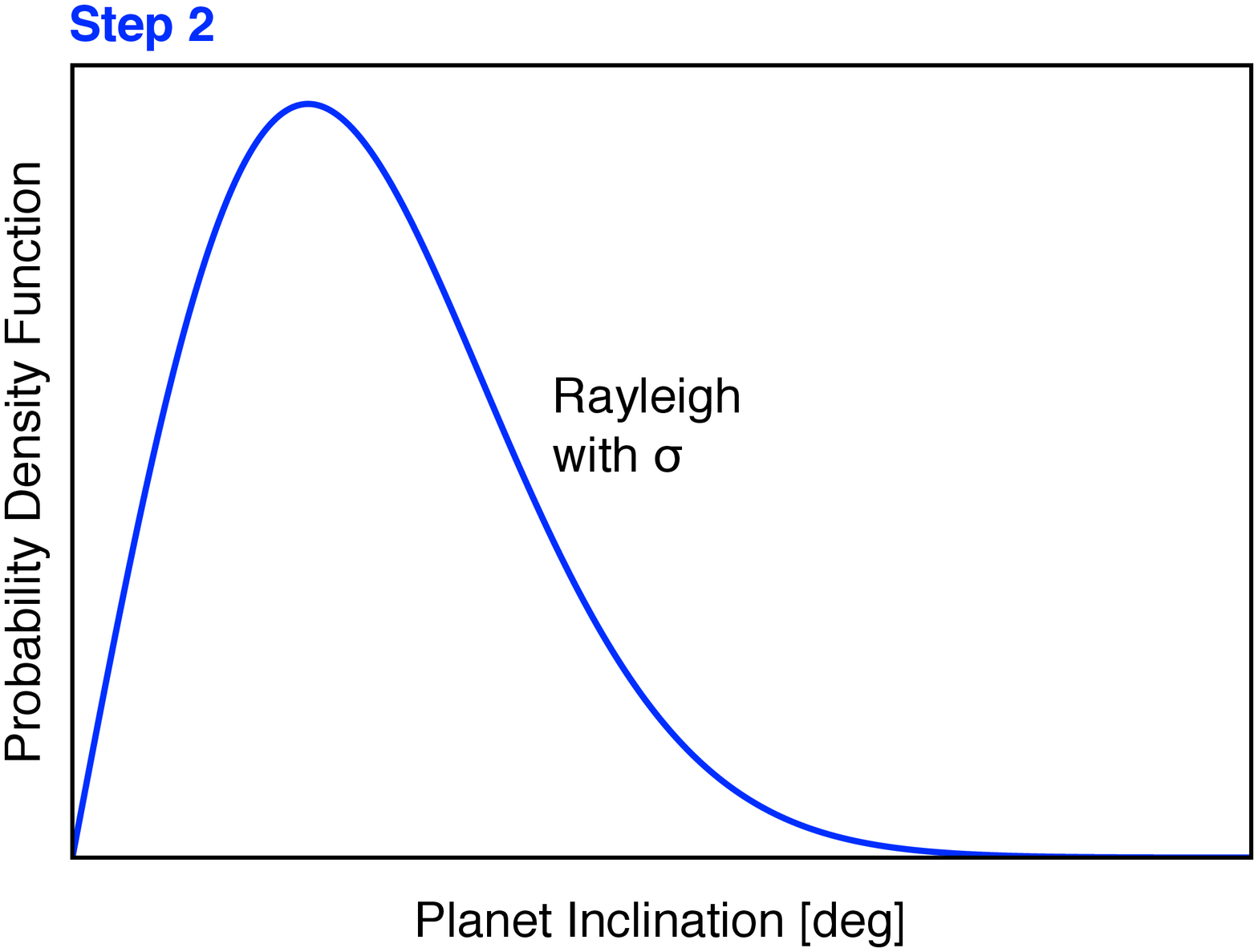}} }
	\vspace{-0.3cm}
	\mbox{ \subfigure{\includegraphics[width=2.8in]{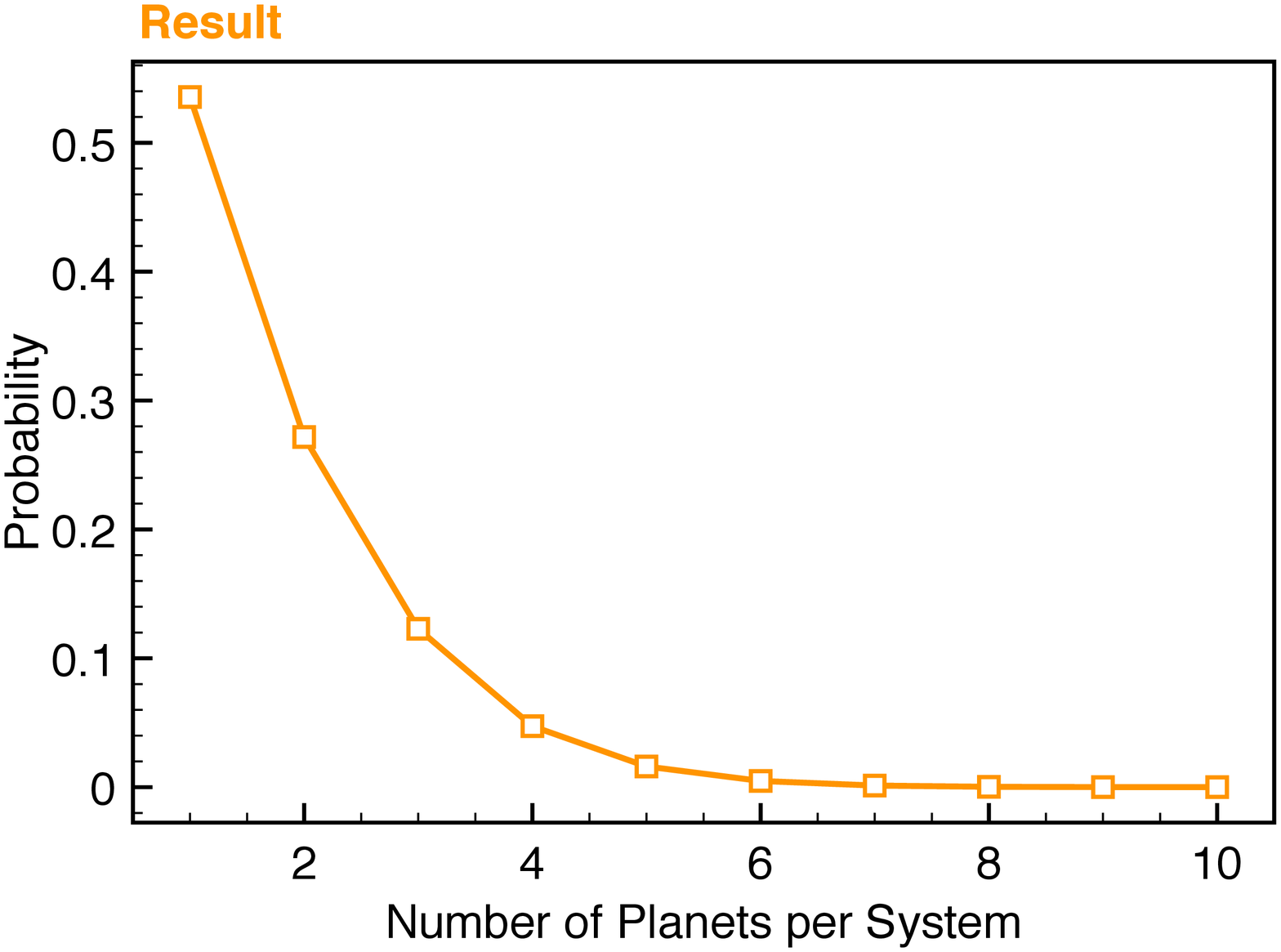}} \quad
	       \subfigure{\includegraphics[width=2.8in]{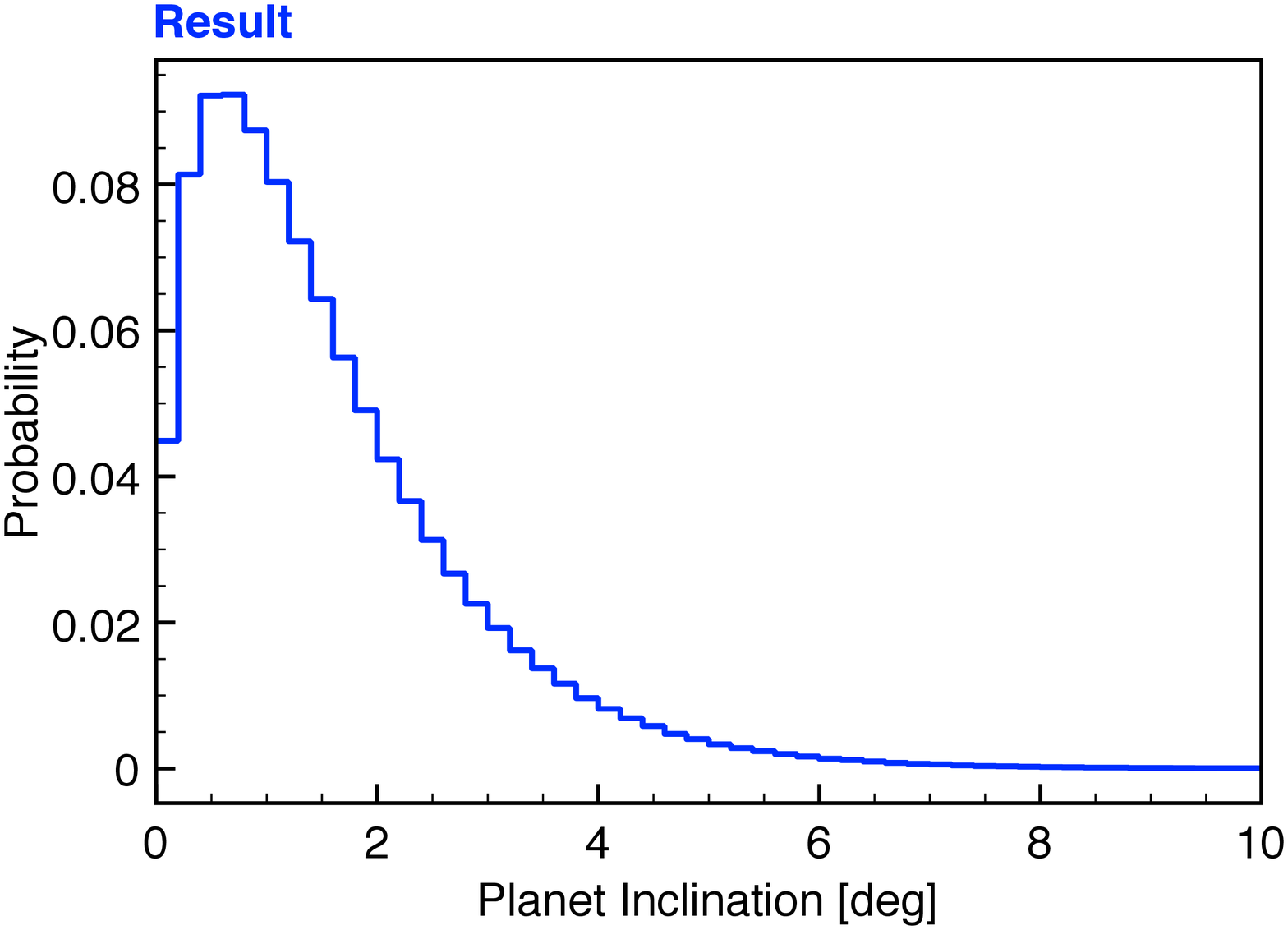}} } 
	\vspace{-0.3cm}
	\mbox{ \subfigure{\includegraphics[width=2.8in]{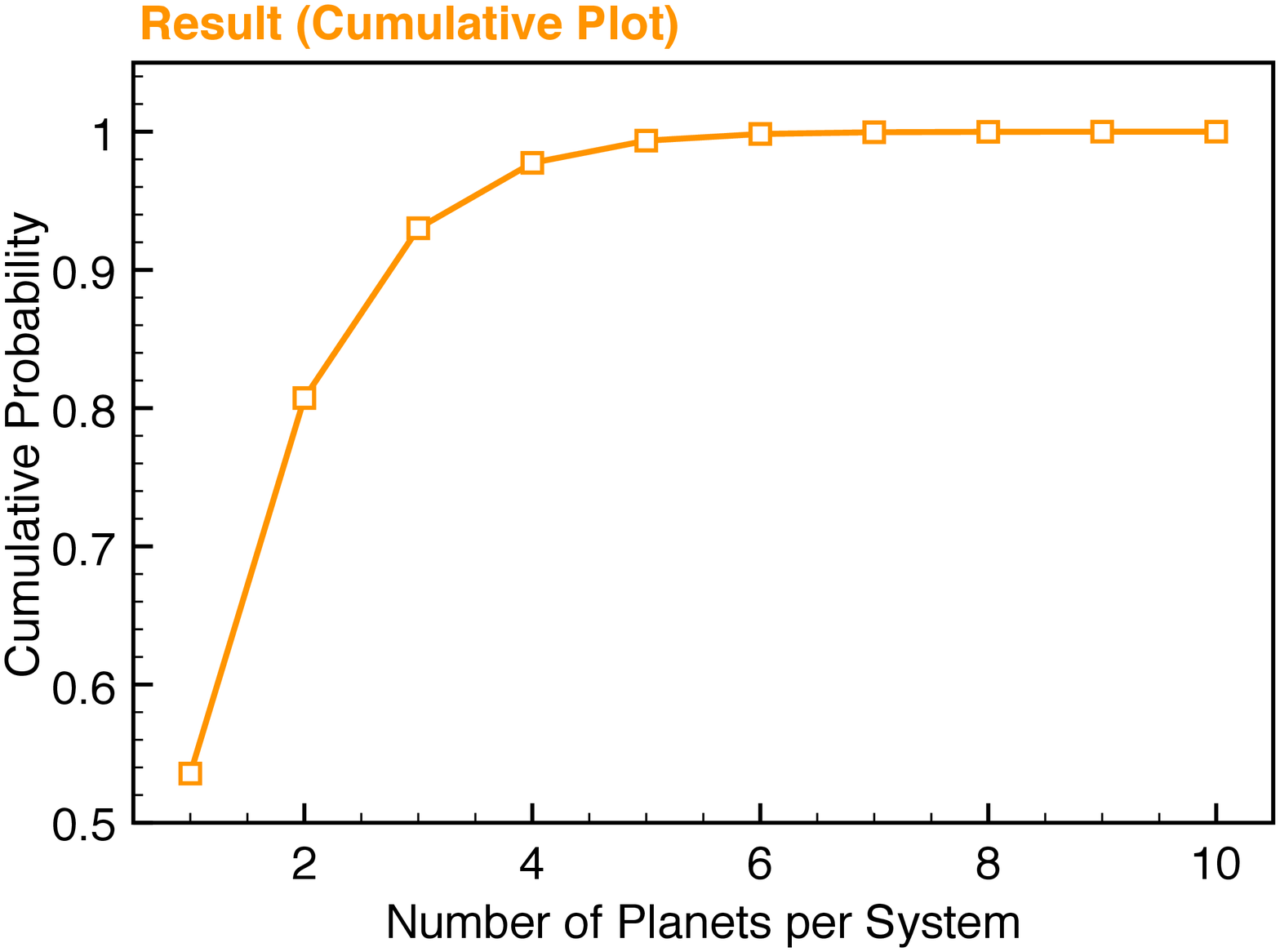}} \quad
	       \subfigure{\includegraphics[width=2.8in]{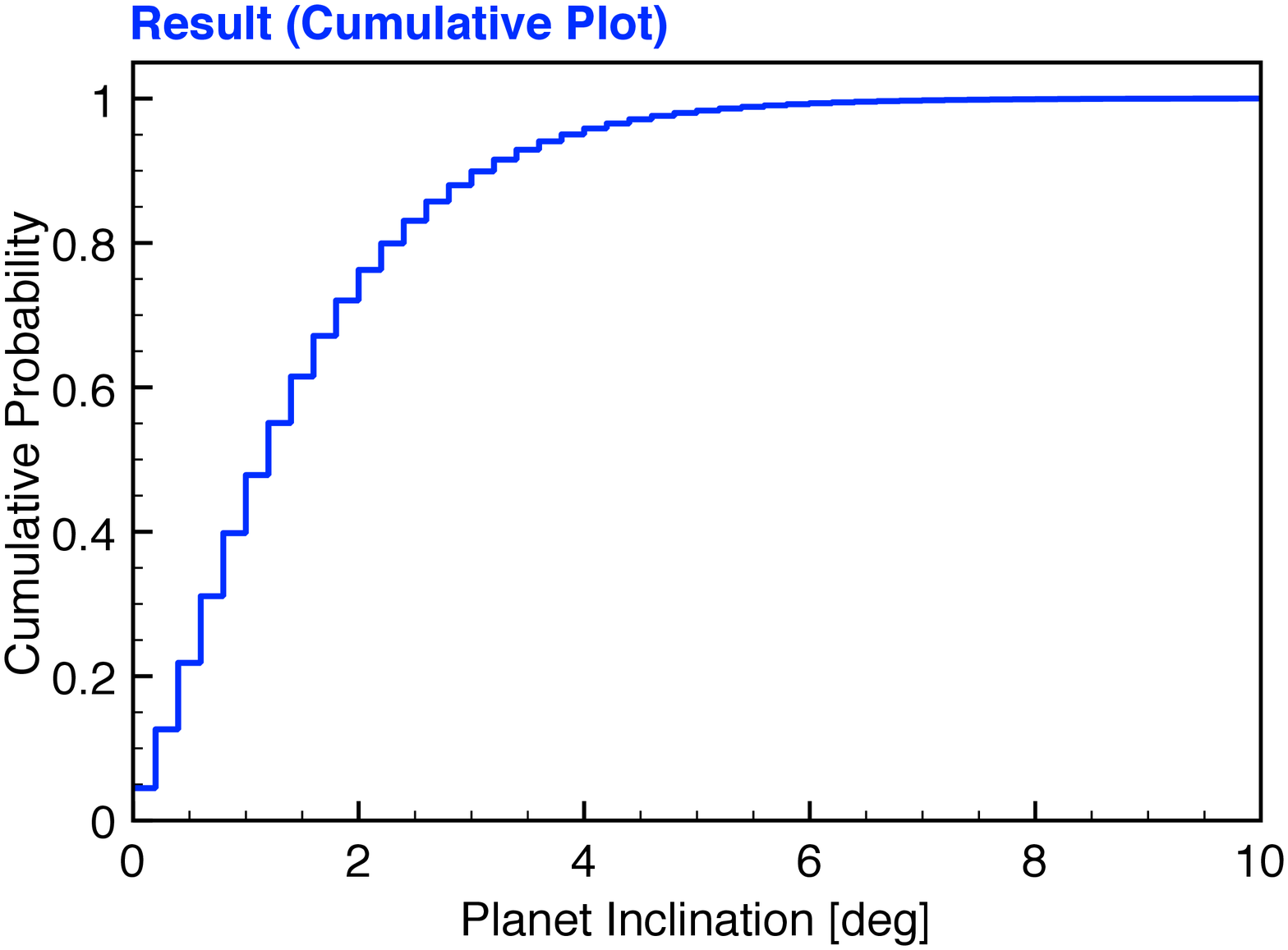}} }
	\caption{One of two best-fit models for multiplicity ({\em left column}) and
	inclination ({\em right column}). The multiplicity distribution is a 
	bounded uniform distribution: for each planetary system (1) we draw 
	a value $N_{\rm max}$ from a modified Poisson distribution with $\lambda=2.25$, 
	and (2) we choose the number of planets by uniformly picking a 
	value between 1 and $N_{\rm max}$. The resulting distributions of 
	multiplicities are shown in the bottom plots.
	The inclination distribution is a Rayleigh of Rayleigh distribution: 
	(1) for each planetary system we draw a value for $\sigma$ from a 
	Rayleigh distribution with parameter $\sigma_{\sigma}=1^{\circ}$, 
	and (2) for each of its planets we draw a value for inclination 
	from a Rayleigh distribution with parameter $\sigma$. The resulting  
	distributions of inclinations are shown in the bottom plots.
\label{truedists}}
\end{figure*}

\begin{figure}[htb]
	\centering
	\includegraphics[width=3.2in]{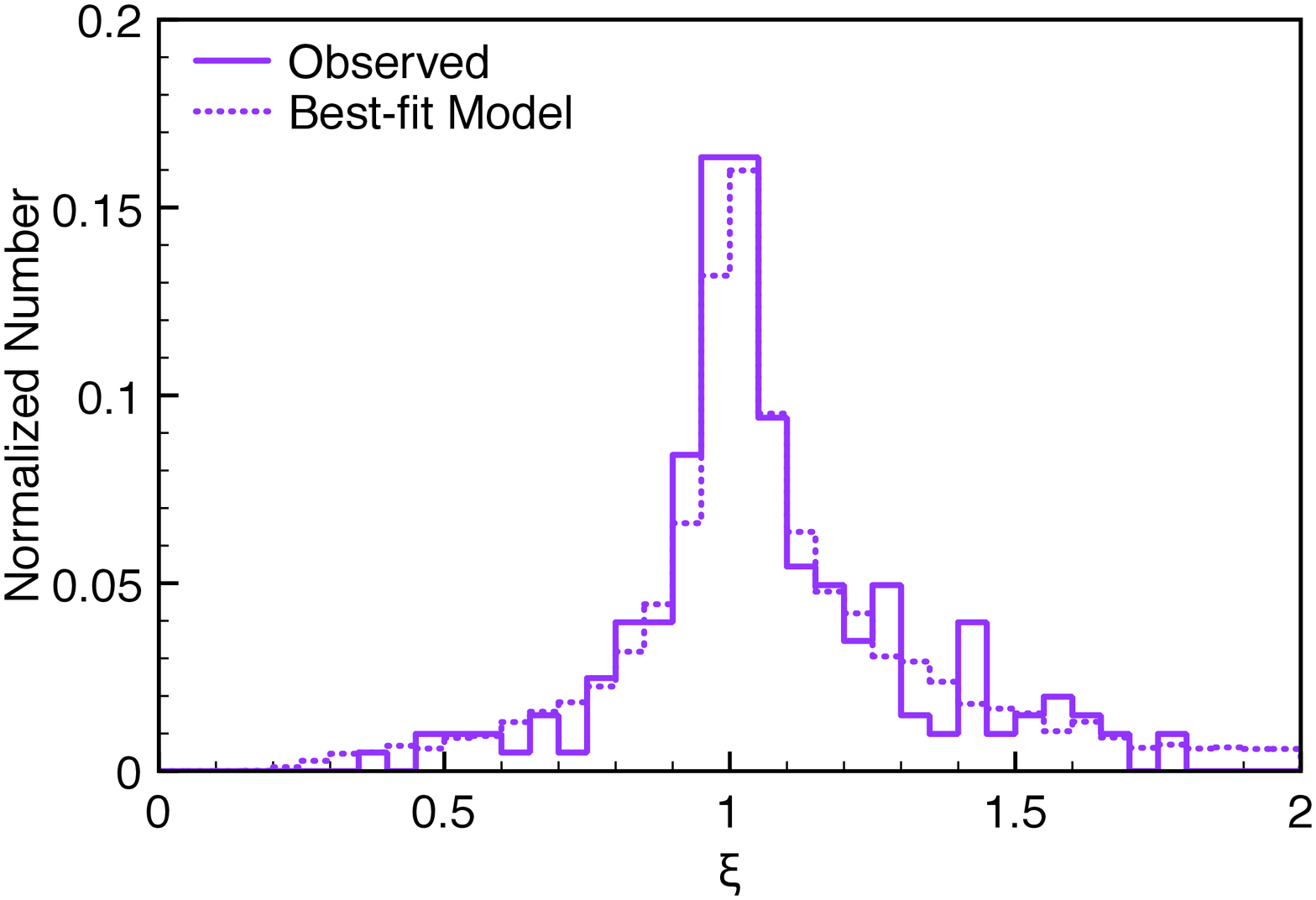}
	\caption{Comparison of the $\xi$ distribution between
          observations and one of two best-fit models,
          here with $\lambda=2.25$ and $\sigma_{\sigma}=1^{\circ}$
          (Section \ref{bestfit_section}).
          The quantity $\xi$ is sensitive to orbital inclinations in multi-planet transiting systems.
          The K-S probability for
          this match is 27.5\%.
\label{xi_comp}}
\end{figure}

We analyze the best-fit model with a bounded uniform ($\lambda=2.25$)
and a Rayleigh of Rayleigh ($\sigma_{\sigma}=1^{\circ}$) distribution
in greater detail. First, in Figure \ref{truedists} we plot the
underlying multiplicity and inclination distributions. The first and 
second rows of the figure show the first and second steps of picking
multiplicity and inclination values from the distributions.  The
plots in the bottom two rows show the resulting histograms of values picked from
these distributions.  Second, we show the comparisons between the
best-fit model and the data.  Table \ref{multfcn_comp} compares the
numbers of observed transiting systems to the numbers of detected
transits from the best-fit model.  There is a reasonable match in each
multiplicity index and an overall match probability of 25.2\%.  Figure
\ref{xi_comp} illustrates the comparison between the distribution of
$\xi$ values from {\em Kepler} observations and the distribution of
$\xi$ values from simulated detections using the best-fit model.  The
numbers in each bin have been normalized to enable comparison, and the
match between distributions has a probability of 27.5\%. Taken
together, the probabilities from each fitted observable ($\chi^2$
probability for the multiplicity fit and K-S probability for the $\xi$
distribution fit) are multiplied to yield a combined probability of
6.9\%.

The low inclinations in our best-fit models have interesting
implications for planet formation and evolution theories. Our findings
that these systems are relatively coplanar (at least out to 200 days)
are in favor of standard models that suggest planet formation within a
protoplanetary disk. In addition, strong influences by external
perturbers such as Kozai processes, scattering, or resonances are not
likely to play a major, lasting role given that these systems do not
generally have high, excited inclinations.
Theories of planet formation and evolution can be tested against the
architecture of planetary systems illuminated by the {\em Kepler}
mission.

%
\def\arraystretch{1.4}
\begin{deluxetable}{lrrrrrrrr}
\tablecolumns{9}
\tablecaption{Multiplicity Vector:\\Number of Systems with $j$ Detectable Planets \label{multfcn_comp}}
\startdata
\hline \hline
\multicolumn{1}{c}{Name} &
\multicolumn{1}{r}{$j=$} &
\multicolumn{1}{c}{1} &
\multicolumn{1}{c}{2} &
\multicolumn{1}{c}{3} &
\multicolumn{1}{c}{4} &
\multicolumn{1}{c}{5} &
\multicolumn{1}{c}{6} &
\multicolumn{1}{c}{7+} \\
\hline
Observed              & & 542   &  85   &  24   &  4   &  1   &  1   &  0 \\
Best-fit 	      & & 540.6 & 92.9  &  19.1 &  3.6 & 0.6  & 0.2  &  0.0 
\enddata
\tablenotetext{}{Comparison of the multiplicity vector between observations and 
one of two best-fit models, here with $\lambda=2.25$ and 
$\sigma_{\sigma}=1^{\circ}$ (see Section \ref{bestfit_section}). The $\chi^2$ 
probability for this match is 25.2\%.
}
\end{deluxetable}

\subsection{Effects of False Positives and Search Incompleteness}

The results of our study do not take into account the existence of
false positives that mimic real planet transit signals in the KOI
dataset.  For the {\em Kepler} sample released by \citet{boru11b},
\citet{mort11} estimated that false positive probabilities were
generally low at $\lesssim10$\%, and an analysis using binary
statistics by \citet{liss12} estimated that $\sim$98\% of the
multi-planet systems are real. For the latest {\em Kepler} sample
released by \citet{bata12}, \citet{fabr12} discussed how $\sim$96\% of
pairs in multi-transiting candidate systems are likely to be real
planets using stability arguments. For the purposes of our study, we
assumed that all planetary candidates are real. We expect that the
existence of false positives can potentially affect the fit between multiplicity
vectors (e.g., Table \ref{multfcn_comp}), particularly for the case of
singly transiting systems where the actual population may actually
have lower numbers than reported due to false positives. We do not
expect that false positives will significantly influence the fit
between normalized transit duration ratios (e.g., Figure
\ref{xi_comp}) since such ratios are only computed for multi-planet
systems, which we expect to be a higher-fidelity sample than singly
transiting systems.

To determine the effect of false positives on our results, we repeated 
the entire analysis described in this paper for two cases. In the 
first case, we removed all false positives described in the 
literature \citep{howa11,bata12,fabr12,sant12,ofir12,colo12}; a total of 
35 false positives were removed from the KOI table. Since we base our 
debiased period and radius distributions on the observed sample, we created 
new model populations and repeated our analyses. In the 
second case, not only did we remove all known false positives but we also 
added simulated false positives to our artificial populations. For each 
population of planetary systems, we added enough false positives (all 
single-planet systems) such that false positives represented 10\% of all 
planets. The properties of the simulated false positives (e.g., radius 
and period) were based on the properties of known false positives.
We then repeated all of our analyses for these populations.
In both cases, we found that our results of intrinsic distributions 
were almost identical in terms of the typical number of planets per system as 
well as their inclinations. Accordingly, we do not expect that the effect 
of false positives will appreciably change our results.

Another effect that can impact our results is {\em Kepler} 
search incompleteness. The {\em Kepler} team is currently
investigating the completeness of the pipeline by inserting artificial
transits into the pipeline and determining their recovery rates.
\citet{bata12} suggested that the KOI catalog released in 2011 by
\citet{boru11b} suffers from incompleteness issues because the
planet gains seen between the 2011 and 2012 catalogs cannot be
completely explained by the longer observation window, and that it is
possible the 2012 catalog still suffers from some incompleteness
issues. This can affect our results in terms of the comparison
between the observations and our model's computed observables, and we
expect that its impact is a function of SNR and other parameters.  
The impact of incompleteness will be better known after the {\em Kepler} team's detailed
study of pipeline completeness is finished, and pipeline completeness
is expected to improve in the future with pipeline upgrades
that are already underway \citep{bata12}.

To investigate the reliability of our results given incompleteness 
issues, we have repeated all simulations and analyses described 
in this paper for higher SNR thresholds of 15, 20, and 25. In all cases, 
we find that our conclusions regarding low intrinsic inclinations are 
the same and that the {\em Kepler} data favors low-inclination 
distributions. However, we find that the best-fit multiplicity distribution 
rises towards larger numbers of planets as we increase to higher SNR 
thresholds. This can be explained as follows. For a given population of 
planetary systems, by imposing higher and higher SNR thresholds on the 
detectability of its transiting planets, planets in two-planet systems are 
more likely to be missed than planets in one-planet systems. This is 
because the outer planet in two-planet systems tends to have a larger 
orbital period than the planet in one-planet systems, and therefore the 
outer planet in two-planet systems will exhibit fewer transits and have 
lower SNR. Consequently, the ratio of observed one-planet to two-planet 
systems should rise with higher SNR thresholds for detectability. This 
effect is seen in our synthetic populations, but is not seen in the observed 
{\em Kepler} data, where the ratio of observed one-planet to two-planet 
systems remains near 6:1 for SNR$\sim$10 to SNR$\sim$25 for the radius 
and period regime considered here. As a result, the best-fit multiplicity 
distribution rises with increasing SNR thresholds in order to match the data.
This curious trend in the data suggests that perhaps there are many 
single-planet systems missing in the KOI table. If this is the case, then 
our multiplicity determinations serve as an upper limit regarding the 
typical number of planets per system. The best way to test this is to 
repeat this analysis in the future when the KOI tables are less biased 
and more complete.

\section{Comparison With the Solar System \\and With Previous Work} \label{discussion}

We compare the results of our study with the properties of planets in
the Solar System as well as with some relevant previous studies.

\subsection{Results Are Consistent With the Solar System}

Based on the underlying multiplicity distribution of our best-fit
models, 75-80\% of planetary systems have 1 or 2 planets with orbital
periods less than 200 days (see Section \ref{bestfit_section}). In
comparison, the Solar System has 1 planet, Mercury, with an orbital
period less than 200 days. Accordingly, our best-fit multiplicity 
distributions are consistent with the Solar System, if we extrapolate 
the results of our study to planetary radii less than 1.5 Earth radii. Note 
that our analysis is agnostic about the number of planets with larger orbital
periods or different radius/mass regimes than considered here; recall
that these limits are due to the parameter space constraints imposed by our study and defined 
in Equations (\ref{stellarcuts}) and (\ref{planetcuts}).  
More data and extension of this study to longer orbital periods and 
smaller planetary radii are warranted before any definitive
multiplicity comparison can be made with the Solar System.

Based on the underlying inclination distributions of our best-fit
models, over 85\% of planets have orbital inclinations less than 3
degrees (see Section \ref{bestfit_section}), suggesting a high degree
of coplanarity. This is compatible with the inclinations seen among
the planets in the Solar System, if we allow ourselves to extrapolate
beyond the period and radii limits of our study. In the Solar System,
7 out of the eight planets (or 87.5\%) have inclinations less than 3
degrees relative to the invariable plane, with Mercury as the
exception.

\subsection{Comparison with Previous Studies}

\citet{liss11} investigated the dynamical properties of multi-planet
systems in the KOI catalog announced by \citet{boru11b}. They also
presented a detailed analysis of the inclinations of {\em Kepler}
planetary systems. To accomplish this, they created a host of
planetary models following different planet multiplicity and
inclination distributions, determined which of those planets were
transiting and detectable, and compared the resulting transit numbers
with the observed transit numbers to determine each model's
goodness-of-fit. Differences between their methods and ours include
different methods for obtaining debiased period and radius
distributions, different radius and period parameter ranges, an
observed $\xi$ distribution that was used as a fitted observable in
our study, and different data sets \citep[we used 
the most recently released KOI data set;][]{bata12}. From their
results, \citet{liss11} ruled out systems with small numbers of
planets and large inclinations as well as systems with large numbers
of planets and small inclinations, which we also found to be
inconsistent with the data (see Figure \ref{statsmap_1}). In addition,
they found a degeneracy in their results between underlying
distributions of inclination and multiplicity; the number of planets
per system could be increased if the inclination was also increased
and still provide good fits to the data, also seen in our Figure
\ref{statsmap_1}. Ultimately they discounted the case of thick systems
with many planets by invoking results from radial velocity surveys
that suggested low inclinations \citep[see discussion in
][]{fabr12}. In our study, we were able to reject that scenario
because it provided poor fits to the observed $\xi$ distribution (see
Figure \ref{statsmap_2}).

\citet{trem12} applied a general statistical model to {\em Kepler} data as
well as to radial velocity surveys to analyze the multiplicity 
and inclination distributions. They concluded that {\em
Kepler} data alone could not place constraints on inclinations, and
acceptable possibilities ranged from thin to even spherical
systems. Systems with large rms inclinations could be fit by the data
as long as some of them had a large number of planets. However, when
jointly analyzing both {\em Kepler} and radial velocity data, they
were able to place constraints on inclinations to show that mean
planetary inclinations are in the range $0-5^{\circ}$. These
relatively low inclinations are consistent with our results showing
how models with lower inclinations provided better fits to the data.

\citet{figu12} determined underlying inclination distributions by
applying information from both HARPS (High Accuracy Radial Velocity
Planet Searcher) and {\em Kepler} surveys. They assumed that if the
HARPS and {\em Kepler} surveys share the same underlying population,
then the different detection sensitivities of these two surveys with
their detected samples of planets should allow determination of the 
underlying inclinations. To do so, they created synthetic populations 
of planets with a given multiplicity distribution as previously
determined using HARPS data, and planets were given various
inclination distributions (aligned, Rayleigh, and isotropic). Each
model's goodness-of-fit was determined by comparing the frequency of
transiting systems with the {\em Kepler} sample. They found that the
best fits were obtained using models with inclinations prescribed by a
Rayleigh distribution with $\sigma \leq 1^{\circ}$.  In our study, we
used a different data set based on the latest release by
\citet{bata12}, we fit for an additional parameter (the intrinsic
multiplicity distribution), and we fit to an extra observable (the
resulting $\xi$ distribution). Given that, our overall results are in
agreement since we also found that low inclinations with a Rayleigh
parameter $\sigma \sim$1$^{\circ}$ are consistent with the data.

\citet{fabr12} presented important properties of multi-planet
candidates in the KOI catalog released by \citet{bata12}. They found
that almost all of these systems are apparently stable (using nominal
mass-radius and circular orbit assumptions), which reinforces the
high-fidelity nature of these multi-planet detections.  In addition,
they derived the mutual inclinations of observed planets using the
distribution of $\xi$ from the {\em Kepler} sample, which we used in
our study as a fitted observable. They found that mutual inclinations
are constrained to the range of 1$-$2.3$^{\circ}$ for observed
planets, and concluded that planetary systems are typically flat.  Our
analysis is different and has some advantages in that we considered 
underlying planet multiplicity and inclination models, ran them
through synthetic observations, and fitted the models' results to
observed transit numbers and $\xi$ distribution.  Our study agrees
with the conclusions of \citet{fabr12} that planetary systems tend to
be relatively coplanar to within a few degrees.

\citet{weis12} implemented an analytical model and investigated an
intrinsic Poisson distribution for planet multiplicity as well as
coplanar and non-coplanar assumptions for inclinations using a
Rayleigh distribution. They found that none of their choices for
multiplicity and inclination distributions produced numbers of
transiting planet systems matching the {\em Kepler} data. This is
generally consistent with our results, where we find that combinations
of Poisson multiplicity distributions and Rayleigh inclination
distributions produce some of the poorest fits. They suggested that
the disagreement between their results and the data could be
potentially explained if planet occurrence is not an independent
process and is instead correlated between planets in the same system,
where the existence of one planet may affect the existence of
additional planets.

\citet{joha12} used a different approach than some previous
authors. They created synthetic triple planet systems by assuming they
could be based off of observed {\em Kepler} triply transiting
systems. They determined which of their synthetic planets could be
transiting and observable, and then calculated the resulting number of
singly, doubly, and triply transiting systems to compare with the
actual number of transits in the {\em Kepler} data. They also repeated
these steps for different inclinations. From fitting to observed
doubly and triply transiting systems, they found low mutual
inclinations of $\lesssim5^{\circ}$.  Although our methods are
different, such low mutual inclinations are compatible with our
findings that these systems are intrinsically thin. In addition,
\citet{joha12} found that numbers of transiting two-planet and
three-planet systems can be matched by underlying three-planet
systems, but they underproduced the number of transiting single-planet
systems. To investigate this further, we have repeated our entire
analysis by considering systems with only one or three planets. We
find that we are not only able to reproduce the numbers of transiting
two-planet and three-planet systems, as did \citet{joha12}, but also
the number of single planet systems.  Of course, by considering only
systems with one and three planets we are not able to match the number
of systems with four transiting planets or greater.

\section{Conclusions} \label{conclusion}

We have investigated the underlying distributions of multiplicity and
inclination of planetary systems by using the sample of planet
candidates discovered by the {\em Kepler} mission during Quarters 1
through 6. Our study included solar-like stars and planets with
orbital periods less than 200 days and with radii of 1.5$-$30
R$_{\Earth}$. We created model populations represented by a total of
two tunable parameters, and we fitted these models to observed numbers
of transiting systems and to normalized transit duration ratios. We
did not include any constraints from radial velocity surveys. Below we
list the main conclusions of our study.

1. From our best-fit models, 75-80\% of planetary systems have 1 or 2 
   planets with orbital periods less than 200 days. This
   represents the unbiased, underlying number of planets per system.

2. From our best-fit models, over 85\% of planets have orbital inclinations 
   less than 3 degrees (relative to a common reference plane), implying a 
   high degree of coplanarity. 

3. Compared to previous work, our results do not suffer from
degeneracies between multiplicity and inclination.  We break the
degeneracy by jointly considering two types of observables that
contain information on both number of planets and inclinations.

4. If we extrapolate down to planet radii less than 1.5 Earth radii,
the underlying multiplicity distribution is consistent with the number
of planets in the Solar System with orbital periods less than 200
days. If we also extrapolate to beyond 200 days, we find that the
underlying distribution of inclinations derived here is compatible
with inclinations in the Solar System.  

5. Our results are also consistent with the standard model of planet
formation in a disk, followed by an evolution that did not have a
significant and lasting impact on orbital inclinations.

Continued observations by the {\em Kepler} mission will improve the
detectability of new candidate planets covering a larger swath of
parameter space, especially to longer orbital periods and smaller
planetary radii.  We anticipate that future statistical work will
further boost our understanding of the underlying architecture of
planetary systems.

\acknowledgments

We thank Dan Fabrycky for useful discussions, as well as the entire {\em Kepler} 
team for procuring such an excellent dataset of planetary systems. We also 
thank the reviewer for helpful comments that improved the paper.

\bibliographystyle{apj}
\bibliography{exostats}

\end{document}